# Incorporating precipitation-related effects on plastic anisotropy of age-hardenable aluminium alloys into crystal plasticity constitutive models

A. Wessel[a,b], E. S. Perdahcıoğlu[c], A. H. van den Boogaard[c], A. Butz[a], W. Volk[b]

[a] Fraunhofer Institute for Mechanics of Materials IWM, Woehlerstrasse 11, 79108 Freiburg, Germany

[b] Chair of Metal Forming and Casting, Technical University of Munich, Walther-Meissner-Strasse 4, 85748 Garching, Germany

[c] Chair of Nonlinear Solid Mechanics, Faculty of Engineering Technology, University of Twente, Drienerlolaan 5, 7522 NB Enschede, The Netherlands

E-mail address: alexander.wessel@iwm.fraunhofer.de

## Abstract

Crystal plasticity finite element simulations are frequently employed to predict the plastic anisotropy of polycrystalline metals based on their crystallographic texture. In age-hardenable aluminium alloys, however, the texture-induced plastic anisotropy is known to affect by precipitation. This paper presents a new modelling approach to incorporate this effect into crystal plasticity constitutive models. The approach focuses on the overall effect of precipitation, which is assumed to result in an additional directional dependency with respect to a global material orientation, superimposed with the texture-induced plastic anisotropy. This additional directional dependency is implemented into a conventional crystal plasticity constitutive model via a modified hardening law that introduces two new parameters, of which only one is treated as a free parameter. To demonstrate the applicability of the new modelling approach, it is applied to an age-hardenable AA6014-T4 aluminium alloy and compared against a state-of-the-art crystal plasticity constitutive model that considers only crystallographic texture. The results demonstrate that the new modelling approach significantly improves the prediction accuracy of the plastic anisotropy for the AA6014-T4 aluminium alloy studied.

## 1. Introduction

The plastic material behaviour of sheet metals is typically anisotropic, i.e. direction dependent. Plastic anisotropy in sheet metals arises during the rolling process, which induces a preferred orientation within the crystallographic texture. This preferred orientation of the crystals, or grains, in the polycrystalline structure leads to direction-dependent plastic material behaviour on the macroscopic scale. To characterise the plastic anisotropy of sheet metals, a wide range of experiments with different loading conditions have been developed over the past decades. Commonly performed experiments for characterising the plastic anisotropy of sheet metals include uniaxial tensile tests in different sheet directions, plane strain tension tests and hydraulic bulge tests, among others. A more comprehensive overview of the experimental characterisation of sheet metals can be found in Banabic et al. (2010, 2020) and Rossi et al. (2022).

An alternative approach to characterising the plastic anisotropy of sheet metals is crystal plasticity simulations, which are also referred to as virtual experiments. Crystal plasticity simulations predict the plastic anisotropy of polycrystalline metals based on their physical origin: the crystallographic texture. To this end, a representative volume element (RVE) of the microstructure, including the crystallographic texture, is generated. By employing a crystal plasticity constitutive model, which describes the plastic deformation behaviour of crystalline materials on the basis of crystallographic slip along the active slip systems, the macroscopic anisotropy of sheet metals can be predicted, or rather virtually characterised. There are many examples in the literature of successful predictions of plastic anisotropy by means of crystal plasticity simulations, particularly for non-heat-treatable aluminium alloys such as the AA1XXX, AA3XXX and AA5XXX series. Inal et al. (2010) used crystal plasticity simulations to predict the normalised yield stresses as well as the r-values (also known as Lankford coefficients) for a 1.0 mm thick continuous cast AA5754 aluminium sheet in the fully recrystallised condition. The results for the normalised yield stresses and the r-values, obtained by means of uniaxial tensile tests at 0°, 15°, 30°, 45°, 60°, 75° and 90° with respect to the rolling direction (RD), were in good agreement with experimental data. Crystal plasticity simulations were also performed by Zhang et al. (2014) for 1.2 mm thick AA3103 aluminium sheets in the H18 and O tempers. For both the AA3103-H18 and AA3103-O aluminium alloys, the experimental results for the normalised yield stresses and the r-values were captured reasonably well by crystal plasticity simulations. Examining an AA1050 aluminium sheet with a thickness of 1.5 mm, Zhang et al. (2015) predicted the normalised yield stresses and the r-values of uniaxial tensile tests in seven different directions with respect to RD with sufficient accuracy.

In Zhang et al. (2016), results of crystal plasticity-based and experimental uniaxial tensile tests in different directions with respect to RD were compared for a cold-rolled AA3104 aluminium alloy. Again, the results of the crystal plasticity simulations were in good agreement with the experimental data. Engler and Aretz (2021) recently performed crystal plasticity simulations to characterise the plastic anisotropy of aluminium sheets made of AA1200-O, AA3104-H19, AA5182-O and AA5050A-H44, with thicknesses ranging from 0.245 to 1.5 mm. Although all four aluminium sheets had distinctly different crystallographic textures, the experimental results for the uniaxial tensile tests and the plane strain tension tests in different directions with respect to RD were reproduced reasonably well by crystal plasticity simulations.

Even though the plastic anisotropy of sheet metals can be predicted accurately by means of crystal plasticity simulations for a wide range of non-heat-treatable aluminium sheets, there is a noticeable deviation in the normalised yield stresses for age-hardenable aluminium alloys, particularly for the AA6XXX series in the T4 temper. For example, Gawad et al. (2015) performed crystal plasticity simulations for AA6016-T4 aluminium sheets with a thickness of 1.0 mm. While the lowest normalised yield stresses were experimentally determined at 45° with respect to RD, crystal plasticity simulations predicted the highest normalised yield stresses in this direction. Simultaneously, the r-values from the crystal plasticity simulations were in good agreement with the experimental data. Hama et al. (2021) reported a similar divergent trend between the normalised yield stresses of experiments and crystal plasticity simulations for AA6022-T4 aluminium sheets with a thickness of 1.0 mm. Also, the crystal plasticity simulations presented in Engler (2022) for 1.2 mm thick sheets made of AA6016-T4 aluminium alloy show a similar divergence in the normalised yield stresses at 45° with respect to RD compared to the experimental data. Further examples of AA6016 aluminium sheets in the T4 temper which exhibited analogous differences between the results of crystal plasticity simulations and experiments are also provided by Habraken et al. (2022). Recently, this deviation between experiments and crystal plasticity simulations was further verified by the authors (Wessel et al., 2023) for a 1.0 mm thick AA6014-T4 aluminium sheet using two different crystal plasticity constitutive models. While the results for the r-values predicted by simulations were in good agreement with the experimental results for both crystal plasticity constitutive models, the results for the normalised yield stresses at 0°, 15°, 30°, 45°, 60°, 75° and 90° with respect to RD showed a diverging trend compared to the experimental data. This means that the experimental normalised yield stresses were lowest at 60° with respect to RD, with the maximum values for the normalised yield stresses determined in RD and the transverse direction (TD). In contrast, crystal plasticity simulations predicted a maximum at around 15° with respect to RD, with local minima in RD and TD.

Furthermore, some literature sources suggest precipitation as a possible explanation for the deviations described in the previous paragraph. In this respect, it is already known that the plastic anisotropy of age-hardenable aluminium alloys can be affected by precipitation. The first reports on precipitation as an additional source of plastic anisotropy were published in the early 1970s for AA2XXX series aluminium alloys. Hosford and Zeisloft (1972) proposed that θ' platelets forming on the {100} crystal planes may affect the texture-induced plastic anisotropy of an Al-4% Cu aluminium alloy. These findings were further supported by Jobson and Roberts (1977), who performed cylindrical cup drawing tests on Al-4% Cu aluminium sheets in different heat treatment conditions. Further examples can be found in Bate et al. (1981, 1982), Hargartner et al. (1998) and Choi et al. (2001), among others. With respect to age-hardenable AA6XXX series aluminium alloys, Kuwabara et al. (2017) analysed the plastic anisotropy of an AA6016 aluminium alloy in the T4 and O temper conditions. The experimental results for the crystallographic texture as well as the r-values were very similar for both tempers. In contrast, the results for the normalised yield stresses were divergent. It was concluded that the plastic anisotropy of AA6016-O was controlled by crystallographic texture alone, while that of AA6016-T4 was governed by both crystallographic texture and Guinier-Preston (GP) zones. Yoshida et al. (2021) analysed an unspecified aluminium alloy from the AA6XXX series in the T4 and O temper conditions. Although the experimental results for the r-values, crystallographic texture and grain size were almost identical, those for the normalised yield stresses were dissimilar. Moreover, crystal plasticity simulations performed for both aluminium alloys were in good agreement with the experimental results for the O temper only. In this study, clusters/GP zones were also mentioned as a possible explanation for the difference in the plastic anisotropy of the AA6XXX series aluminium alloy in the T4 temper.

This study presents a new modelling approach to incorporate precipitation-related effects on the plastic anisotropy of age-hardenable aluminium alloys into crystal plasticity constitutive models. The new approach is implemented into a conventional crystal plasticity constitutive model that originally predicts plastic anisotropy based on crystallographic texture as the only mechanism. The extended crystal plasticity constitutive model, which accounts for both texture-induced plastic anisotropy as well as precipitation-related effects, is then applied to an AA6014-T4 aluminium alloy to predict the plastic anisotropy of uniaxial tensile tests in different directions with respect to RD. The results of the crystal plasticity simulations are verified by comparing them with experimental results for the AA6014 aluminium alloy in both T4 and O temper conditions.

## 2. Crystal plasticity modelling of precipitation-related effects

### 2.1 Initial crystal plasticity framework

The crystal plasticity constitutive model used in this study is the rate-independent model introduced by Aşık et al. (2020). This model describes plastic deformation based solely on crystallographic slip as the only deformation mechanism. Additional deformation mechanisms, such as twinning or transformation-induced plasticity, are not taken into account. Full details of the crystal plasticity constitutive model and its implementation can be found in Perdahcıoğlu et al. (2018) and Perdahcıoğlu (2024). In the following, a summary of the kinematics, flow rule and hardening formulations used in the model is provided.

The total deformation is described by the deformation gradient **F**, which can be multiplicatively decomposed into an elastic and a plastic part:

$$\mathbf{F} = \mathbf{F}_{\mathrm{e}}\mathbf{F}_{\mathrm{p}}. \tag{1}$$

Here, $\mathbf{F}_{\mathrm{e}}$ represents the reversible elastic deformation resulting from stretching and rotating the crystal lattice, while $\mathbf{F}_{\mathrm{p}}$ describes the irreversible plastic deformation due to crystallographic slip. The elastic deformation gradient pertains to the intermediate configuration, whereas plastic deformations are described in the current configuration. Following Eq. (1), the velocity gradient **L** is formulated as:

$$\mathbf{L} = \dot{\mathbf{F}}\mathbf{F}^{-1} = \left(\dot{\mathbf{F}}_{\mathrm{e}}\mathbf{F}_{\mathrm{p}} + \mathbf{F}_{\mathrm{e}}\dot{\mathbf{F}}_{\mathrm{p}}\right)\mathbf{F}_{\mathrm{p}}^{-1}\mathbf{F}_{\mathrm{e}}^{-1} = \mathbf{L}_{\mathrm{e}} + \mathbf{F}_{\mathrm{e}}\mathbf{L}_{\mathrm{p}}\mathbf{F}_{\mathrm{e}}^{-1}. \tag{2}$$

The plastic part of the velocity gradient $\mathbf{L}_{\mathrm{p}}$ acting in the current configuration is, by definition, the sum of the shear rates $\dot{\gamma}^{\alpha}$ acting on each slip system $\alpha$ (Mandel, 1965; Rice, 1971):

$$\mathbf{L}_{\mathrm{p}} = \sum_{\alpha=1}^{n} \dot{\gamma}^{\alpha}\mathbf{m}^{\alpha} \otimes \mathbf{n}^{\alpha}. \tag{3}$$

The unit vectors $\mathbf{m}^{\alpha}$ and $\mathbf{n}^{\alpha}$ represent the slip direction and slip plane normal, respectively. The parameter $n$ denotes the total number of slip systems. For face-centred cubic (FCC) materials, such as aluminium alloys, 12 slip systems, crystallographically denoted as $\{111\}\ \langle 1\bar{1}0\rangle$, are incorporated into the crystal plasticity constitutive model. The flow rule is expressed in a rate-independent form. In this context, crystallographic slip occurs only when the resolved shear stress $\tau^{\alpha}$ of a slip system $\alpha$ equals the slip resistance $\tau_{\mathrm{crit}}^{\alpha}$. Mathematically, this is expressed as

$$\phi^{\alpha} = \tau^{\alpha} - \tau_{\mathrm{crit}}^{\alpha} \leq 0. \tag{4}$$

Based on the principle of maximum dissipation, Eq. (4) is solved as an optimisation problem using the interior point method. This method inherently resolves the ambiguity of slip systems arising from the linear dependence that occurs when more than five slip systems become active, see Taylor (1934), as does the rate-dependent formulation. The resolved shear stress $\tau^\alpha$ of a slip system $\alpha$ in Eq. (4) is computed from the Cauchy stress tensor $\boldsymbol{\sigma}$, the slip direction $\mathbf{m}^\alpha$ and the slip plane normal $\mathbf{n}^\alpha$, resulting in the following equation:

$$\tau^\alpha = \boldsymbol{\sigma} : (\mathbf{m}^\alpha \otimes \mathbf{n}^\alpha). \tag{5}$$

The hardening rule for the slip resistance $\tau^\alpha_{\text{crit}}$ in Eq. (4) is based on a physics-based hardening law. Accordingly, the following Taylor-type hardening model (Taylor, 1934) is employed:

$$\tau^\alpha_{\text{crit}} = \tau + \mu b \sqrt{\textstyle\sum_\beta q^{\alpha\beta} \rho^\beta}\ . \tag{6}$$

The parameters $\tau$, $\mu$ and $b$ are the lattice friction, the shear modulus and the length of the Burgers vector, respectively. The total dislocation density of a slip system is governed by a linear ordinary differential equation

$$\dot{\rho}^\beta = \frac{\dot{\gamma}^\beta}{\gamma_s} (\rho_s - \rho^\beta). \tag{7}$$

Here, $\gamma_s$ represents a parameter controlling the saturation rate, while $\rho_s$ is the saturation dislocation density. In the initial state, the total dislocation density $\rho^\beta$ is defined by the initial dislocation density $\rho_0$. All material-dependent parameters $\tau$, $\mu$, $b$, $\rho_s$, $\rho_0$ and $\gamma_s$ of the crystal plasticity constitutive model are assumed to be identical for each slip system. Interaction between two distinct slip systems $\alpha$ and $\beta$ is incorporated via a physics-based interaction matrix. Following Franciosi and Zaoui (1982), Kubin et al. (2008) and Khadyko et al. (2016), the interaction matrix distinguishes between six types of dislocation interactions for face-centred cubic materials: self ($q_0$), coplanar ($q_1$), collinear ($q_2$), orthogonal ($q_3$), gissile ($q_4$) and sessile ($q_5$). Thus, the interaction matrix is given in the following form:

$$
q^{\alpha\beta} = \begin{pmatrix}
q_0 & q_1 & q_1 & q_3 & q_5 & q_4 & q_3 & q_4 & q_5 & q_2 & q_4 & q_4 \\
 & q_0 & q_1 & q_5 & q_3 & q_4 & q_4 & q_2 & q_4 & q_4 & q_3 & q_5 \\
 & & q_0 & q_4 & q_4 & q_2 & q_5 & q_4 & q_3 & q_4 & q_5 & q_3 \\
 & & & q_0 & q_1 & q_1 & q_2 & q_4 & q_4 & q_3 & q_4 & q_5 \\
 & & & & q_0 & q_1 & q_4 & q_3 & q_5 & q_4 & q_2 & q_4 \\
 & & & & & q_0 & q_4 & q_5 & q_3 & q_5 & q_4 & q_3 \\
 & & & & & & q_0 & q_1 & q_1 & q_3 & q_5 & q_4 \\
 & & & & & & & q_0 & q_1 & q_5 & q_3 & q_4 \\
 & & & sym. & & & & & q_0 & q_4 & q_4 & q_2 \\
 & & & & & & & & & q_0 & q_1 & q_1 \\
 & & & & & & & & & & q_0 & q_1 \\
 & & & & & & & & & & & q_0
\end{pmatrix}. \tag{8}
$$

Slip systems in the interaction matrix in Eq. (8) are ordered as follows: $(111)[\bar{1}10]$, $(111)[\bar{1}01]$, $(111)[0\bar{1}1]$, $(\bar{1}11)[110]$, $(\bar{1}11)[101]$, $(\bar{1}11)[0\bar{1}1]$, $(1\bar{1}1)[110]$, $(1\bar{1}1)[\bar{1}01]$, $(1\bar{1}1)[011]$, $(\bar{1}\bar{1}1)[\bar{1}10]$, $(\bar{1}\bar{1}1)[101]$ and $(\bar{1}\bar{1}1)[011]$.

## 2.2 Incorporating precipitation-related effects

Magnesium (Mg) and silicon (Si) are the principal alloying elements of the AA6XXX series aluminium alloys. These aluminium alloys are age-hardenable, and thus different types of precipitates form during the ageing process from the supersaturated solid solution state (SSSS), depending on the ageing time and temperature. The general precipitation sequence for age-hardenable AA6XXX series aluminium alloys is as follows (Dutta and Allen, 1991; Edwards et al., 1998; Marioara et al., 2005):

SSSS → Mg-/Si-clusters → Guinier–Preston (GP) zones → β″ → β′, U1, U2, B′ → β, Si

In the initial stages of solidification, atomic clusters with varying numbers of Mg and Si atoms start to form. These clusters gradually transform into GP zones and further into β″-precipitates. The latter precipitate is associated with peak-aged conditions and has a needle-like shape aligned along the <100> crystal directions. Upon over-ageing, i.e. when the aluminium alloy exhibits a decrease in strength and hardness, several types of particles form, including β′-precipitates. In the equilibrium state, β-precipitates as well as Si-particles are present. For the T4 temper (i.e. solution heat treated and naturally aged), atomic clusters have been repeatedly reported in the literature (Murayama et al., 1998; Pogatscher et al., 2011; Marceau et al., 2013; Zandbergen et al., 2015a; Zandbergen et al., 2015b; Aruga et al. 2015). Moreover, Murayama et al. (2001) also observed GP zones in the T4 temper condition for an AA6XXX series aluminium alloy. Also, Ghosh et al. (2021) mentioned cluster and GP zones in the context of AA6XXX series aluminium alloys in the T4 temper. In addition to the type of precipitation, the shape of these clusters and the GP zones can also vary. Results from Fallah et al. (2015) revealed that clusters initially form with a spherical shape and later elongate along the <100> crystal directions. A study by Zandbergen et al. (2015) also indicated that their shape can range from spherical to needle-like.

Although different types of precipitation with varying shapes have been reported in the literature for AA6XXX series aluminium alloys in the T4 temper, there is evidence suggesting that precipitation affects the texture-induced plastic anisotropy of age-hardenable aluminium alloys. As mentioned in the introduction, Hosford and Zeisloft (1972) proposed that precipitation affects the texture-induced plastic anisotropy of an Al-4% Cu aluminium alloy. Their results were further supported by Bate et al. (1981, 1982), Hargartner et al. (1998) and Choi et al. (2001). The recent reports by Kuwabara et al. (2017) and Yoshida et al. (2021) provide further evidence that the plastic anisotropy of age-hardenable aluminium alloys, and in particular AA6XXX series aluminium alloys in the T4 temper, is affected by precipitation. As the exact physical mechanisms behind this phenomenon remain unclear, the new modelling approach for incorporating precipitation-related effects into crystal plasticity constitutive models focuses on the overall effect of precipitation on texture-induced plastic anisotropy. Rather than providing a complete physical explanation at the microstructure level, the new modelling approach aims to capture this effect in the simplest possible way. In this context, the approach is based on the hypothesis that the overall effect of precipitation, i.e. second-phase particles, results in an additional directional dependency with respect to a global material orientation, which superimposes with the texture-induced plastic anisotropy. With respect to the microstructure, this kind of effect might be caused by ellipsoidal cluster/GP zones being oriented in a global preferred direction and/or spherical cluster/GP zones being distributed with respect to a global preferred direction as shown in Fig. 1, among other possibilities. Again, a detailed explanation of the actual causes and microstructural mechanisms is beyond the scope of this study.

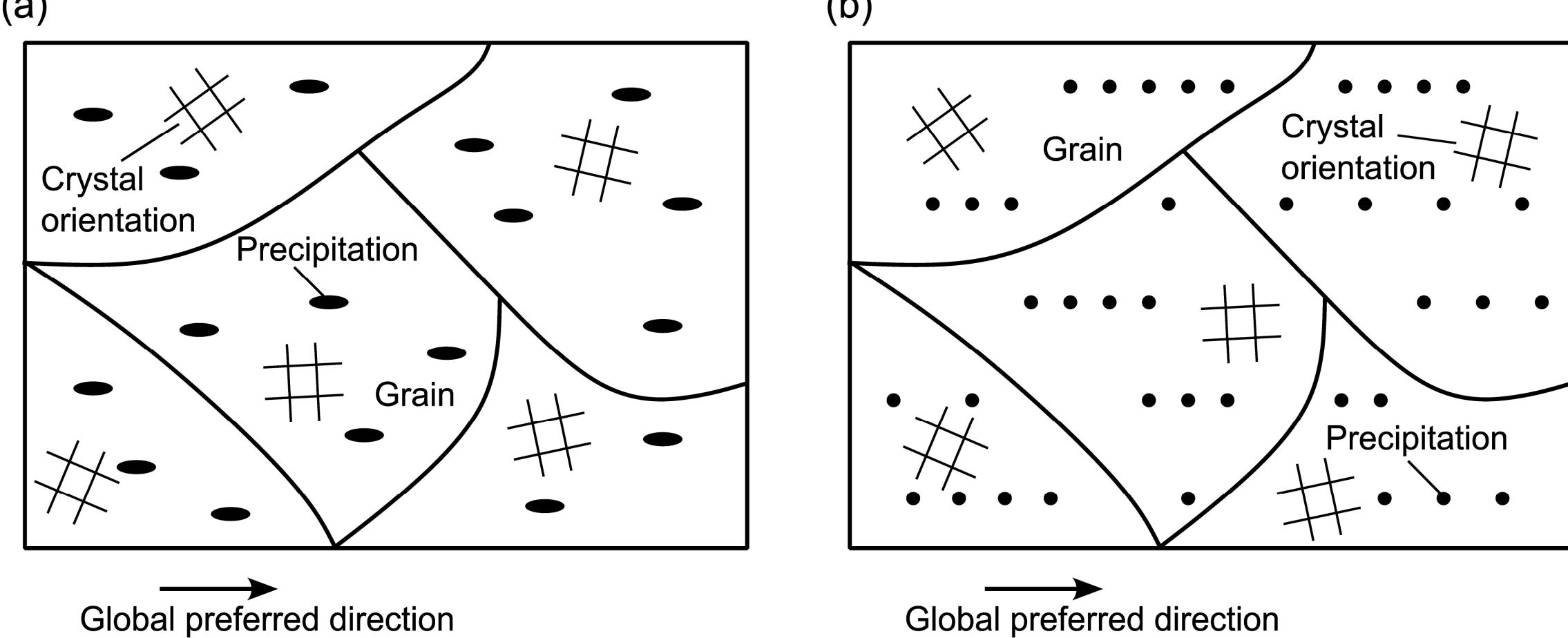

Fig. 1: Illustration of the underlying hypothesis for the new modelling approach: precipitation induces an additional directional dependency with respect to the global material orientation due to (a) ellipsoidal shapes that are oriented in a preferred direction or (b) spherical precipitates that are distributed in a preferred direction, among other possibilities.

Assuming that the effect of precipitation in AA6XXX series aluminium alloys in the T4 temper induces an additional directional dependency with respect to a global material orientation, or rather a global preferred direction, it can be further concluded that this directional dependency is linked to crystallographic slip and thus can be incorporated into crystal plasticity constitutive models. To do so, the slip resistance $\tau^{\alpha}_{\mathrm{crit}}$ of the initial crystal plasticity constitutive model in Eq. (6) - initially identical for all slip systems - is redefined as:

$$\tau^{\alpha}_{\mathrm{crit}} = \tau_{\mathrm{ini}} + \tau^{\alpha}_{\mathrm{ppt}}. \tag{9}$$

The initial lattice friction $\tau_{\mathrm{ini}}$ corresponds to the lattice friction as originally defined in Eq. (6), while $\tau^{\alpha}_{\mathrm{ppt}}$ represents the lattice friction due to precipitation. The latter is a slip-system-dependent variable, which is expressed as

$$\tau^{\alpha}_{\mathrm{ppt}} = \left(\frac{s^{\alpha}}{0.5} \cdot \tau_{\mathrm{ppt,max}}\right). \tag{10}$$

The variable $\tau_{\mathrm{ppt,max}}$ represents the maximum lattice friction due to precipitation, whose magnitude is scaled by a Schmid-like factor $s^{\alpha}$. Unlike the original Schmid factor, which is calculated with respect to the loading direction, the Schmid-like factor $s^{\alpha}$ adjusts the effect of the maximum lattice friction $\tau_{\mathrm{ppt,max}}$ on a slip system $\alpha$, depending on its orientation with respect to the global preferred direction **d**. To this end, the Schmid-like factor $s^{\alpha}$ is calculated as:

$$s^{\alpha} = \cos(\phi) cos(\lambda). \tag{11}$$

Whereas $\phi$ denotes the angle between the slip plane normal $\mathbf{n}^{\alpha}$ and the global preferred direction $\mathbf{d}$, $\lambda$ characterises the angle between the slip direction $\mathbf{m}^{\alpha}$ and the global preferred direction $\mathbf{d}$.

Although the physical mechanisms of precipitation on the texture-induced plastic anisotropy of age-hardenable aluminium alloys in the T4 temper are not considered, their overall effect is reduced to the direction of influence with respect to a global material orientation and its magnitude. Within the extended crystal plasticity constitutive model, this is modelled as a constant change in slip resistance by introducing two parameters: the global preferred direction $\mathbf{d}$ and the maximum lattice friction due to precipitation $\tau_{\mathrm{ppt,max}}$. Any secondary effects of dislocation multiplication are not taken into account. This approach was chosen first to keep the modelling as simple as possible, and second to demonstrate the concept in the absence of experimental observation or evidence to support alternative claims.

## 3. Application to AA6014 aluminium alloy

### 3.1 Materials and temper conditions

The age-hardenable aluminium sheets studied in this work are made from AA6014-T4 aluminium alloy (Trademark Advanz™ 6F - e170) and have a thickness of 1.0 mm. The chemical composition, according to the supplier Novelis Switzerland SA, is summarised in Table 1.

Table 1: Nominal chemical composition of AA6014-T4 aluminium sheets given in wt.% as declared by Novelis Switzerland SA.

| Si | Fe | Cu | Mn | Mg | Cr | Zn | Ti | V | Others, each | Others, total |
|---|---|---|---|---|---|---|---|---|---|---|
| 0.30 - 0.60 | ≤ 0.35 | ≤ 0.25 | 0.05 - 0.20 | 0.40 - 0.80 | ≤ 0.20 | ≤ 0.15 | ≤ 0.10 | 0.05 - 0.20 | ≤ 0.05 | ≤ 0.15 |

To verify the effect of temper conditions on the plastic anisotropy of age-hardenable aluminium alloys, as shown in Kuwabara et al. (2017) for an AA6016 aluminium alloy, specimens were annealed from the initial T4 temper to the O temper, following the ASM guidelines (ASM Handbook, Volume 4: Heat Treating). Thus, specimens made of AA6014-T4 aluminium alloy were heated to an annealing temperature of 415°C at a constant heating rate of 50°C/h in a Nabertherm N11/H chamber. After being kept at this temperature for 2 h, the specimens were cooled at a rate of ≤ 30°C/h.

### 3.2 Experimental characterisation

The experimental characterisation of the AA6014 aluminium alloy in the T4 and O temper conditions consisted of electron backscatter diffraction (EBSD) measurements and uniaxial tensile tests conducted in seven directions. The EBSD measurements were performed on the longitudinal cross-section using a Zeiss Sigma 300 scanning electron microscope (SEM) equipped with an EBSD system from EDAX. Data recording was carried out with APEX™ software and data processing was performed using OIM Analysis™ 8.6. For both the AA6014-T4 and AA6014-O specimens, the area of the EBSD measurements was approximately 2.5 mm × 0.9 mm. Scans were conducted using a hexagonal grid with a step size of 3.0 μm and an accelerating voltage of 20 kV. Post-processing of the EBSD data was carried out using the MATLAB toolbox MTEX 5.7.0 (Bachmann et al., 2010). Following the recommendations of Field (1997), only measurement points with a confidence index greater than 0.1 were considered in the post-processing. A misorientation of 5° was used for grain reconstruction and only grains composed of more than ten measurement points were counted in the analysis.

The mechanical behaviour of the AA6014 aluminium alloy in the T4 and O tempers was characterised by uniaxial tensile tests at 0°, 15°, 30°, 45°, 60°, 75° and 90° with respect to RD. All tests were performed on a ZwickRoell Kappa 50 DS uniaxial testing machine at a constant engineering strain rate of 0.002 $s^{-1}$. During testing, two tactile extensometers were used to measure the change in gauge length in both the longitudinal and transverse directions of the specimen. Specimens were manufactured by water jet cutting and had a gauge length of 80 mm and a gauge width of 20 mm, in accordance with DIN EN ISO 6892. Three specimens were tested for each direction for the AA6014-T4 aluminium alloy, while two specimens were examined for AA6014-O. Results of the experimental characterisation have already been partially presented in Wessel et al. (2023).

### 3.3 Crystal plasticity simulations

The free software package Neper 3.5.2 (Quey, 2011) was utilised to generate a cube-shaped RVE for AA6014-T4 with a normalised edge length of 1.0. Each of the 1000 grains in the RVE was assigned one crystallographic orientation, which was derived from a reconstruction of the experimentally measured orientation density function (ODF) of the AA6014-T4 aluminium alloy using the MATLAB toolbox MTEX 5.7.0. The RVE was discretised by 40x40x40 linear hexahedral elements with full integration (element C3D8 in Abaqus). To define periodic boundary conditions, the procedure outlined by Schmidt (2011) was applied. Crystal plasticity simulations, using both the initial crystal plasticity framework in Section 2.1 and the new modelling approach in Section 2.2, were performed with the commercial finite element software Abaqus/Standard 2021. The hardening parameters of both crystal plasticity constitutive models were identified by a reverse engineering approach. For the initial crystal plasticity constitutive model, the hardening parameters were adjusted to match the experimental stress-strain curve at 0° with respect to RD using the commercial optimisation software LS-OPT 6.0. For the new modelling approach, the identification of the hardening parameters followed a two-step process. As with the initial crystal plasticity constitutive model, the hardening parameters $\tau_{\mathrm{ini}}$, $\rho_0$, $\rho_{\mathrm{s}}$ and $\gamma_{\mathrm{s}}$ were first fitted to the experimental stress-strain curve at 0° with respect to RD using LS-OPT 6.0. Subsequently, the hardening parameters $\tau_{\mathrm{ini}}$ and $\tau_{\mathrm{ppt}}$ were manually adjusted to match the normalised yield stresses at 15°, 30°, 45°, 60°, 75° and 90° with respect to RD. These simulations were performed based on a texture rotation, where the RVE was loaded in RD with the texture rotated by the respective angle. Errors associated with this approach are considered acceptable when grains of the RVE are globular in shape, as later shown in Section 4.1.

The following four studies were performed to assess the new modelling approach regarding its suitability in predicting the plastic anisotropy of age-hardenable aluminium alloys in the T4 temper:

1. Crystal plasticity simulations of uniaxial tensile tests at 0°, 15°, 30°, 45°, 60°, 75° and 90° with respect to RD using the initial crystal plasticity constitutive model were conducted based on a texture rotation. These simulations serve as a reference for the new modelling approach.
2. Using the new modelling approach, a parameter study was carried out to analyse the effect of the two new parameters on the normalised yield stresses and on the r-values. For the global preferred direction **d,** the following angles $\zeta$ with respect to RD were considered: 0°, 15°, 30°, 45°, 60°, 75° and 90°. The maximum lattice friction due to precipitation $\tau_{\mathrm{ppt,max}}$ was set to ±10 MPa. For each configuration, i.e. global preferred direction and maximum lattice friction due to precipitation, seven uniaxial tensile tests at 0°, 15°, 30°, 45°, 60°, 75° and 90° with respect to RD were simulated by rotating the texture. The results of the normalised yield stresses and r-values were then compared with the experimental data for the AA6014-T4 aluminium alloy.
3. Based on the results of the parameter study, the effect of the new modelling approach on the entire yield surface was analysed for the most suitable configuration. To this end, a total of 112 crystal plasticity simulations were performed using both the initial crystal plasticity constitutive model and the new modelling approach. Different stress states or rather proportional strain paths were defined under plane stress condition following the procedure outlined by Butz et al. (2019). Thus, proportional strain paths were systematically prescribed using two angles $\varphi$ and $\theta$. The resulting stress-strain curves of all crystal plasticity simulations were evaluated using a specific plastic work per unit volume of 15.49 MPa, corresponding to a uniaxial true plastic strain of 0.08 in RD. The yield points obtained were then used to identify the parameters of the Yld2004-18p yield function (Barlat et al., 2005) via the least squares method. Since two parameters of the Yld2004-18p yield function are known to be dependent, the parameters $c'_{12}$ and $c'_{13}$ were set to unity, as suggested by van den Boogaard et al. (2016). Additionally, parameters of the Yld2004-18p associated with out-of-plane anisotropy were set to unity, corresponding to the isotropic value.

4. To analyse the new modelling approach excluding the effect of texture-induced plastic anisotropy, it was applied to a microstructure model with uniformly distributed crystal orientations. To this end, 1000 Euler angles were randomly drawn from a uniform ODF generated with the MATLAB toolbox MTEX 5.7.0. Since this procedure is not unique, five sets of 1000 Euler angles were analysed. For each of the five sets, or rather microstructure models, uniaxial tensile tests at 0°, 15°, 30°, 45°, 60°, 75° and 90° with respect to RD were performed by utilising the initial crystal plasticity constitutive model and the new modelling approach. Again, uniaxial tensile tests in different directions were performed based on a texture rotation.

## 4. Results

### 4.1 Crystallographic texture

The experimentally measured inverse pole figure (IPF) maps for the AA6014 aluminium alloy in the T4 and O tempers are illustrated in Fig. 2. Each IPF map contains more than 5000 globular grains with an average size of approximately 412 μm$^2$ and 370 μm$^2$ for the AA6014-T4 and AA6014-O aluminium alloys, respectively. For both tempers, <001> crystal axes of the grains are preferably oriented perpendicular to the normal direction of the sheet metal.

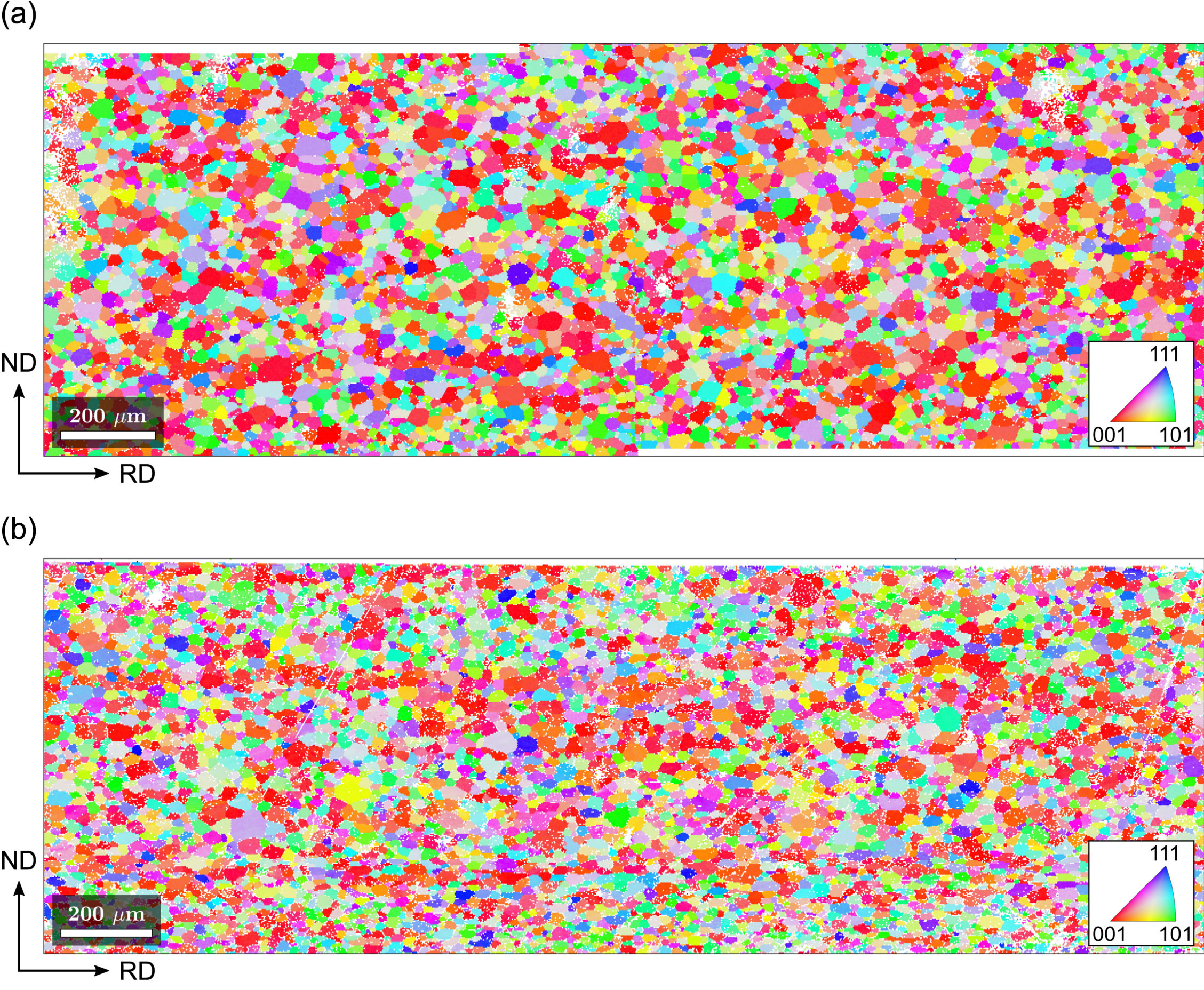

Fig. 2: Inverse pole figure (IPF) maps of the longitudinal cross-section for (a) AA6014-T4 and (b) AA6014-O. IPFs were plotted with respect to the normal direction (ND). Only measurement points with a confidence index greater than 0.1 are illustrated.

The ODFs for AA6014-T4 and AA6014-O in Fig. 3 are nearly identical. Areas with high intensities are associated with a cube texture component, which is crystallographically denoted as {001} <100>. The maximum intensity for the AA6014 aluminium alloy in the T4 and O tempers is 13 and 12 multiples of a random density (MRD), respectively.

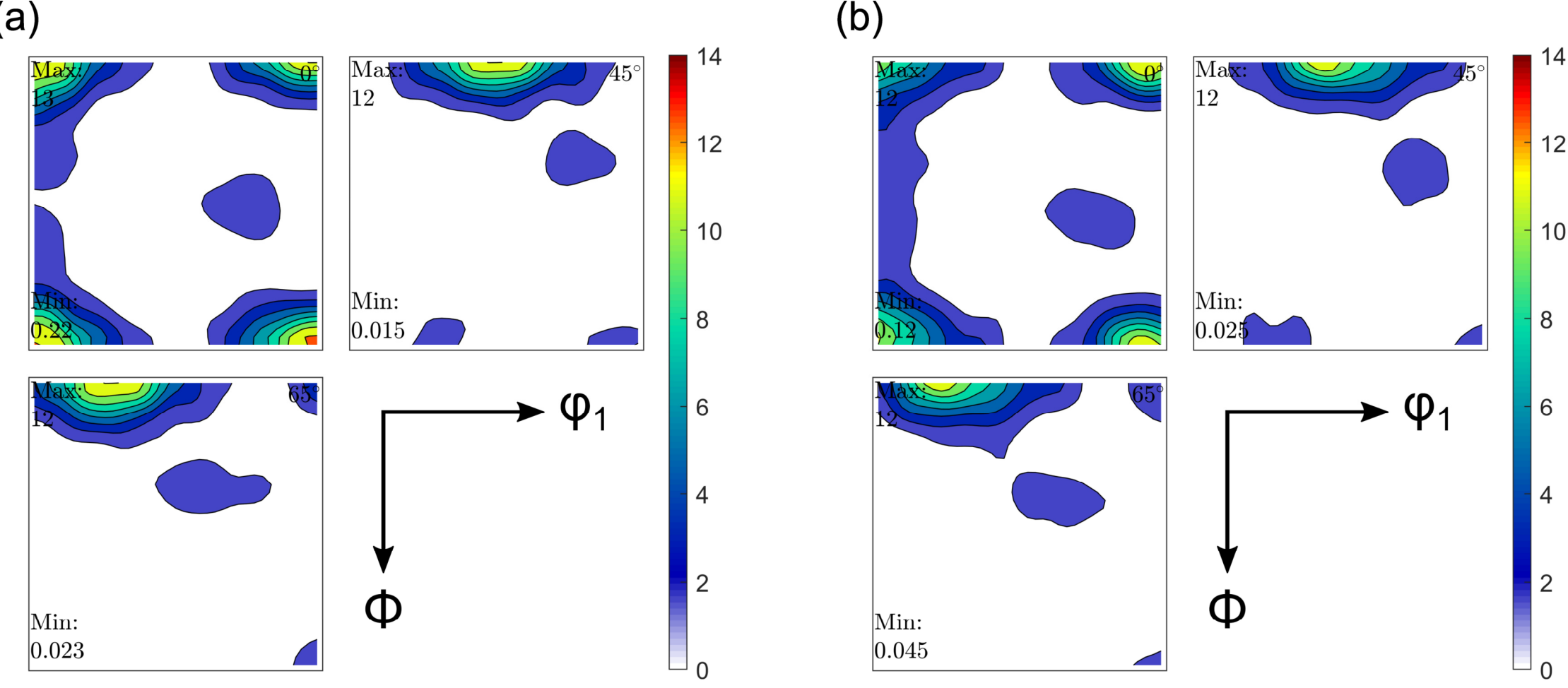

Fig. 3: Orientation density functions (ODF) for (a) AA6014-T4 and (b) AA6014-O in the longitudinal cross-section shown as $\varphi_2$-sections at 0°, 45° and 65° through the reduced Euler space. Only measurement points with a confidence index greater than 0.1 are taken into account for ODF computation.

## 4.2 Mechanical properties

Stress-strain curves of the uniaxial tensile tests at 0°, 15°, 30°, 45°, 60°, 75° and 90° with respect to RD in Fig. 4 highlight significant differences in the hardening behaviour of the AA6014 aluminium alloy in the T4 and O tempers. While the yield strength of the AA6014-T4 aluminium alloy varies from 234.1 to 243.7 MPa, the maximum yield strength of AA6014-O is less than half this and amounts to 112.8 MPa. Furthermore, the tempers exhibit differences in their plastic anisotropy. The stress-strain curves for AA6014-T4 are highest in RD, whereas the highest stress-strain curve for AA6014-O occurs at 30° with respect to RD. Additionally, the stress-strain curves for AA6014-O show a distinctly serrated behaviour.

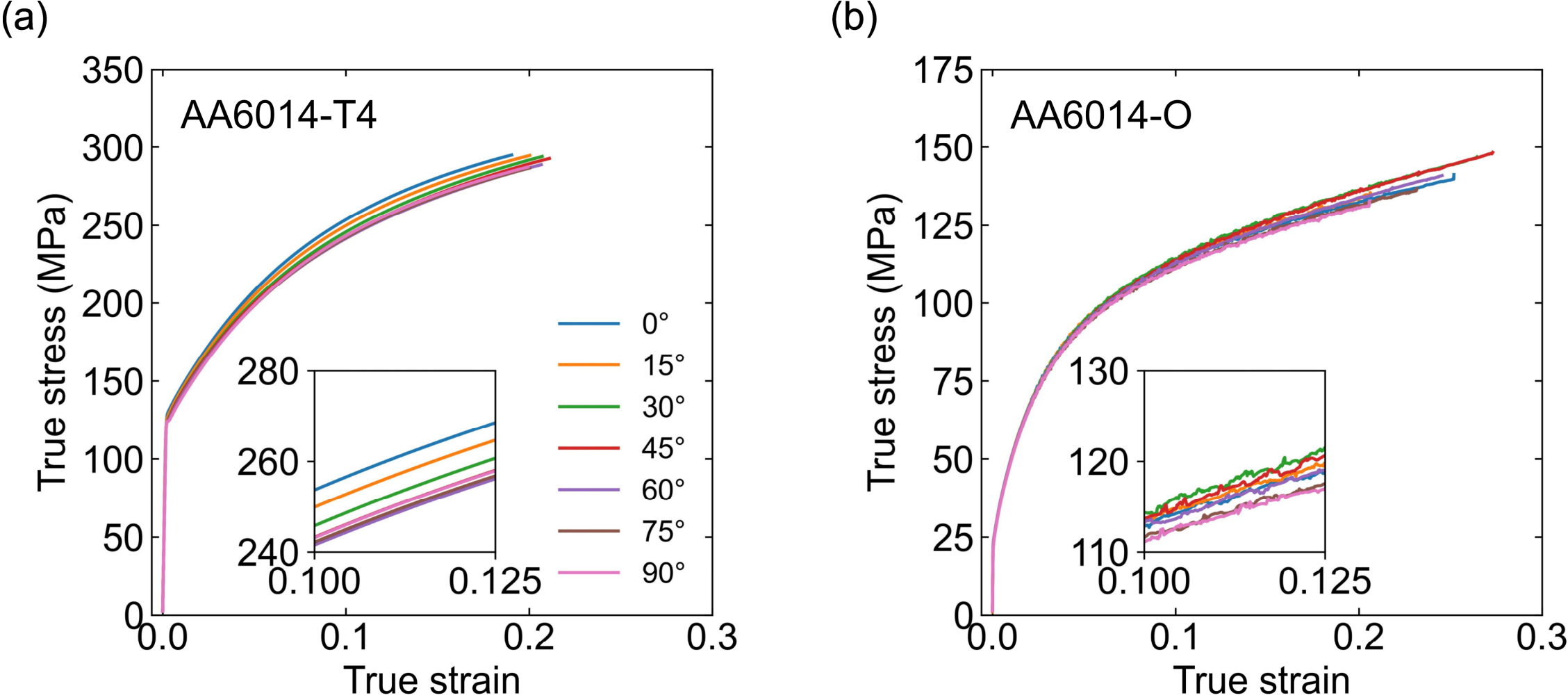

Fig. 4: Stress-strain curves at 0°, 15°, 30°, 45°, 60°, 75° and 90° with respect to RD for (a) AA6014-T4 and (b) AA6014-O. Only one repetition is illustrated as a representative example.

Differences in plastic anisotropy are also evident in Fig. 5, which shows the normalised yield stresses determined at 15.49 and 14.72 MPa specific plastic work for the AA6014-T4 and the AA6014-O aluminium alloys, respectively. This corresponds to a true plastic strain of 0.08 for AA6014-T4 and 0.15 for AA6014-O, and these values were chosen to ensure that the plastic anisotropy remains constant. When comparing the development of the normalised yield stresses for different directions with respect to RD, a complementary curve progression can be observed. At the same time, the results for the r-values are relatively similar for both aluminium alloys. The r-values and other mechanical properties obtained from the uniaxial tensile tests are summarised in Table 2.

Table 2: Mechanical properties of AA6014 aluminium alloy in the T4 and O temper conditions as obtained by uniaxial tensile tests in different directions with respect to RD. Values represent the arithmetic mean of the three and two repetitions for the AA6014-T4 and AA6014-O aluminium alloys, respectively.

| Material | Direction | Flow stress[a] (MPa) | Yield strength[b] (MPa) | Uniform elong.[b] (%) | r-value[c] (-) |
|---|---|---|---|---|---|
| AA6014-T4 | 0° | 240.5 | 242.4 | 20.4 | 0.81 |
| | 15° | 237.7 | 240.1 | 21.3 | 0.70 |
| | 30° | 235.0 | 238.5 | 22.2 | 0.56 |
| | 45° | 232.7 | 236.5 | 22.9 | 0.50 |
| | 60° | 231.6 | 234.4 | 23.5 | 0.52 |
| | 75° | 232.7 | 234.7 | 21.9 | 0.69 |
| | 90° | 233.5 | 235.3 | 21.6 | 0.79 |
| AA6014-O | 0° | 124.6 | 109.6 | 24.9 | 0.72 |
| | 15° | 125.2 | 110.1 | 23.1 | 0.66 |
| | 30° | 126.6 | 112.7 | 30.1 | 0.52 |
| | 45° | 126.0 | 112.8 | 31.2 | 0.46 |
| | 60° | 124.8 | 110.1 | 27.3 | 0.50 |
| | 75° | 123.6 | 108.3 | 24.1 | 0.62 |
| | 90° | 122.5 | 108.1 | 23.3 | 0.73 |

[a] True stress at a specific plastic work of 15.49 MPa for AA6014-T4 and 14.72 MPa for AA6014-O (corresponds to a true plastic strain of 0.08 and 0.15, respectively)

[b] Engineering value

[c] Analysed between 0.1 and 0.175 true plastic strain

4.3 Reference results

Figure 5 illustrates the normalised yield stresses and the r-values at 0°, 15°, 30°, 45°, 60°, 75° and 90° with respect to RD as obtained for the AA6014-T4 aluminium alloy using the initial crystal plasticity constitutive model. The normalised yield stresses as obtained from the crystal plasticity simulations in Fig. 5 (a) are highest at 30° with respect to RD and exhibit a rather complementary curve progression compared to the experimental results for AA6014-T4. Conversely, the normalised yield stresses as predicted by the initial crystal plasticity constitutive model correlate well with the experimental results for AA6014-O. For the r-values, the results of the initial crystal plasticity constitutive model are in good agreement with the experimental results for AA6014-T4 and AA6014-O, which are very similar. In the subsequent analysis, the results of the initial crystal plasticity constitutive model will serve as a reference for the outcomes of the new modelling approach.

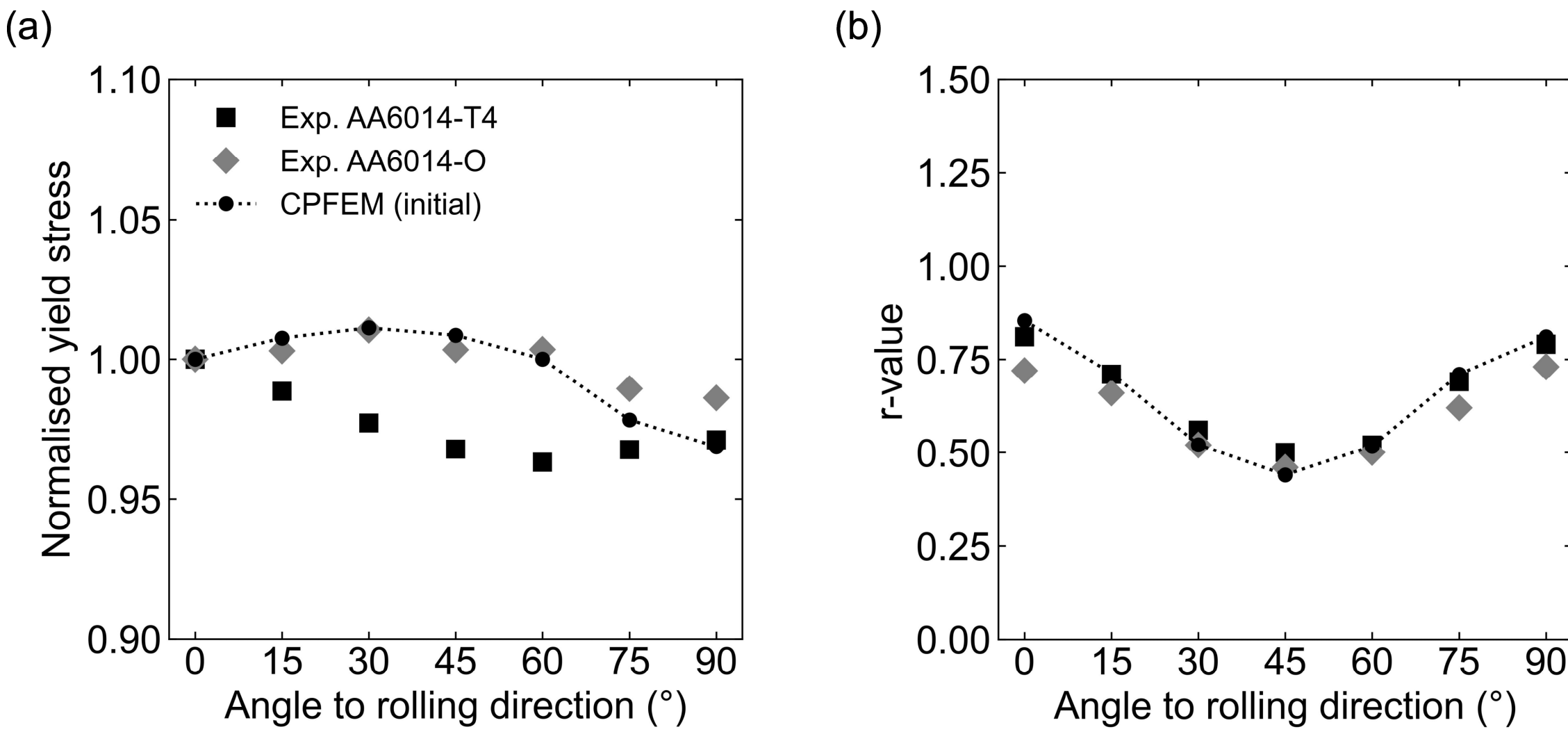

Fig. 5: (a) Normalised yield stresses and (b) r-values predicted by the initial crystal plasticity constitutive model compared with experimental results for AA6014-T4 and AA6014-O. In line with the experimental data, yield stresses of the initial crystal plasticity constitutive model were determined using a specific plastic work of 15.49 MPa, which corresponds to a true plastic strain of 0.08. r-values were analysed within the range of 0.1 to 0.175 true plastic strain.

4.4 Parameter study

To analyse the new modelling approach in more detail, a parameter study was conducted to investigate the effect of the two new parameters, i.e. the global preferred direction **d** and the maximum lattice friction due to precipitation $\tau_{\mathrm{ppt,max}}$. While an angle $\zeta$ at 0°, 15°, 30°, 45°, 60°, 75° and 90° with respect to RD was used for the global preferred direction **d**, the variable $\tau_{\mathrm{ppt,max}}$ was set to ±10 MPa. The results in Fig. 6 demonstrate that the new modelling approach affects the normalised yield stresses and that different curve progressions can be modelled depending on the global preferred direction or angle considered. Overall, the best match between the experimental yield stresses of the AA6014-T4 aluminium alloy and the new modelling approach, or rather the extended crystal plasticity constitutive model, is given for a global preferred direction of 45° with respect to RD and $\tau_{\mathrm{ppt,max}} = -10$ MPa in Fig. 6 (d).

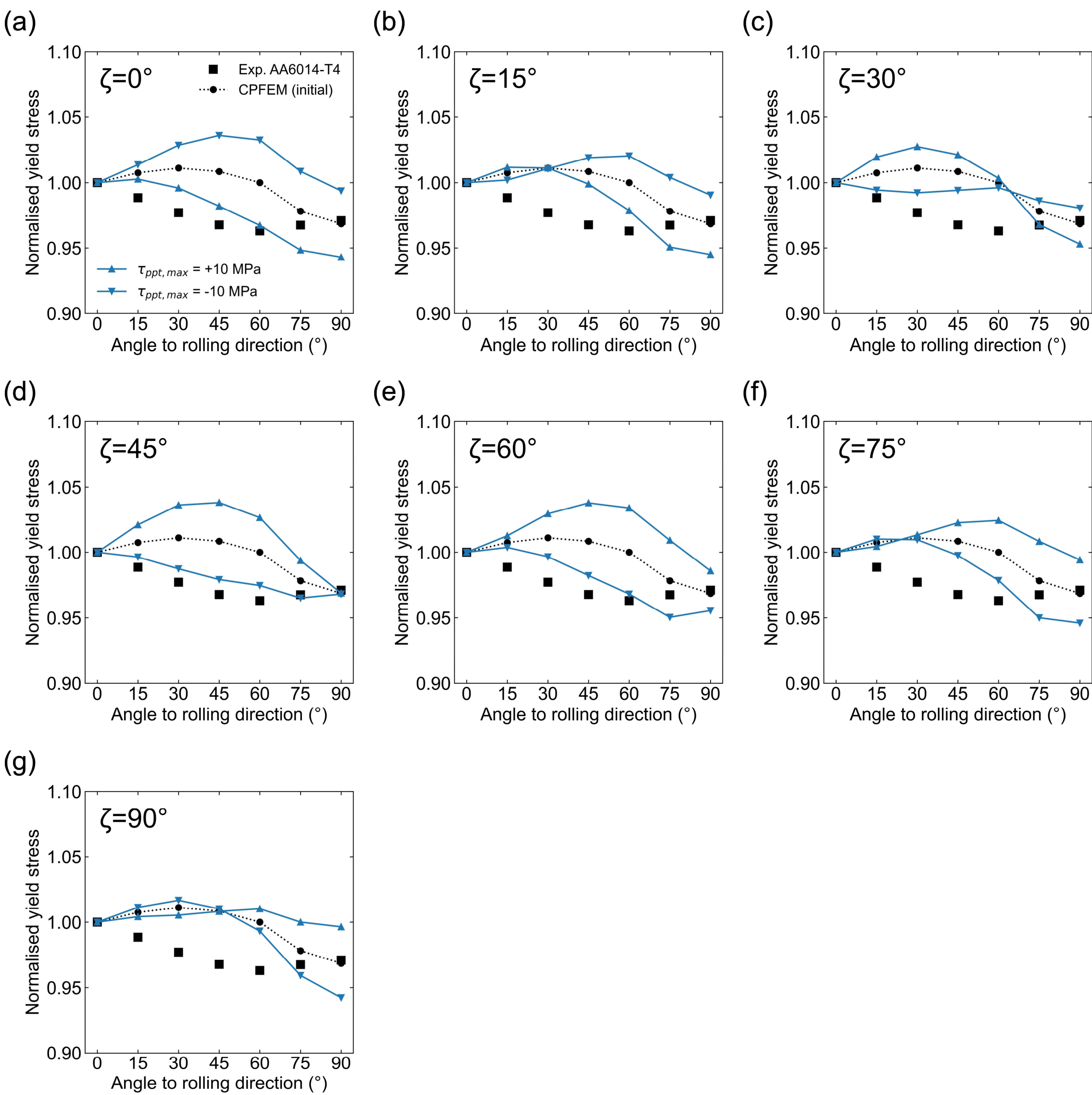

Fig. 6: Normalised yield stresses obtained by the extended crystal plasticity constitutive model using an angle $\zeta$ at (a) 0°, (b) 15°, (c) 30°, (d) 45°, (e) 60°, (f) 75° and (g) 90° with respect to RD as a global preferred direction. Experimental results for AA6014-T4 as well as the results of the initial crystal plasticity constitutive model are also shown by way of comparison. All yield stresses were determined using a specific plastic work of 15.49 MPa.

The effect of the new modelling approach on the results for the r-values in Fig. 7 is less pronounced than on the results for the normalised yield stresses in Fig. 6. Differences are greatest at a global preferred direction of 0°, 15°, 75° and 90° with respect to RD. The results of the new modelling approach utilising a global preferred direction of 30°, 45° and 60° with respect to RD are nearly identical to those of the initial crystal plasticity model, which serves as a reference. As was the case in respect of the results for the normalised yield stresses in Fig. 6, the extended crystal plasticity constitutive model using a global preferred direction of 45° with respect to RD and $\tau_{ppt,max} = -10$ MPa shows the best agreement with the experimental r-values of the AA6014-T4 aluminium alloy in Fig. 7 (d).

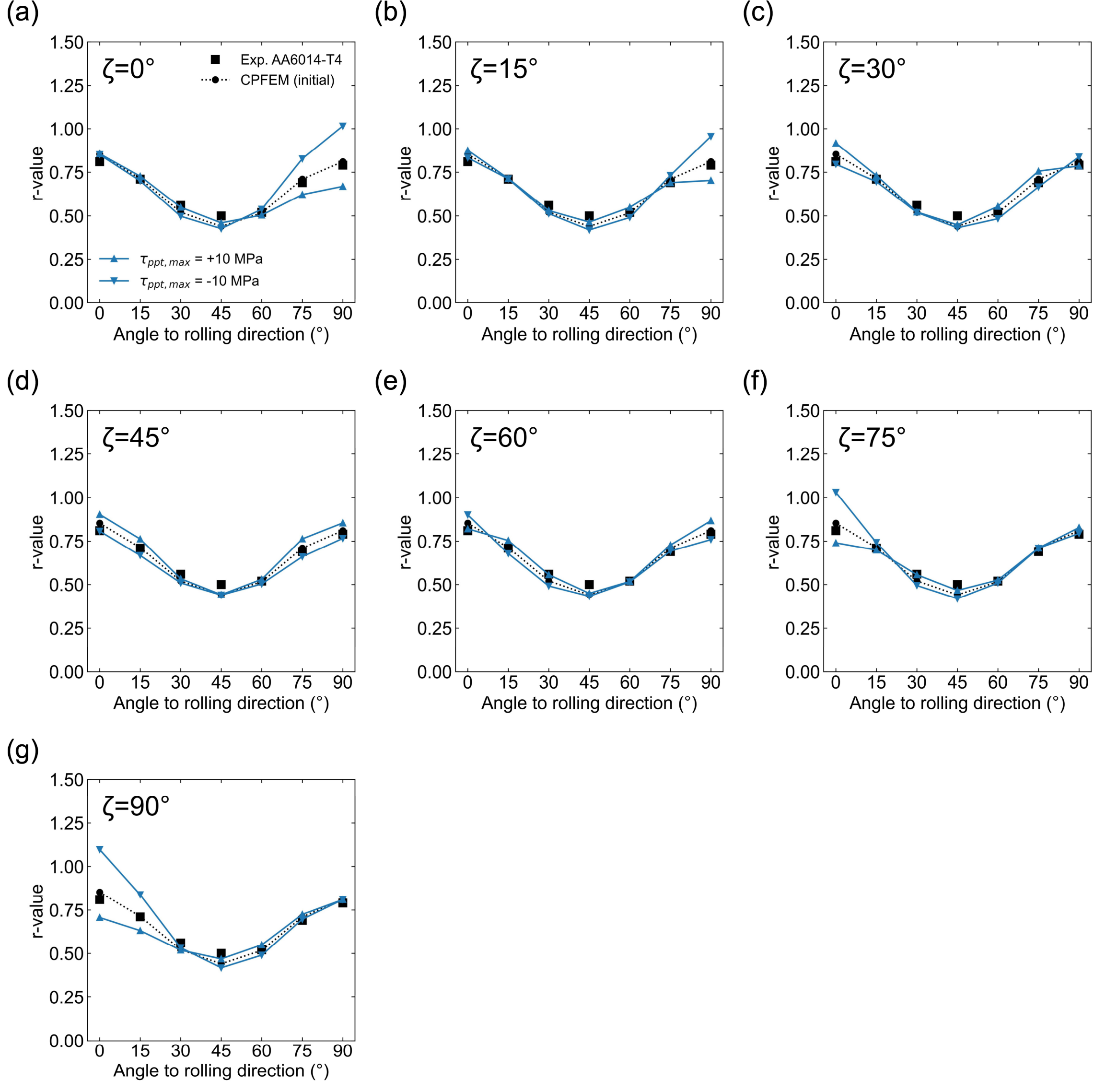

Fig. 7: r-values as obtained by the extended crystal plasticity model considering an angle $\zeta$ at (a) 0°, (b) 15°, (c) 30°, (d) 45°, (e) 60°, (f) 75° and (g) 90° with respect to RD as a global preferred direction. Experimental results for AA6014-T4 as well as the results of the initial crystal plasticity model are also shown by way of comparison. All r-values were analysed between 0.1 and 0.175 true plastic strain.

4.5 Extended crystal plasticity constitutive model

Since the results of the new modelling approach or rather the extended crystal plasticity constitutive model which considered a global preferred direction $\mathbf{d}$ at 45° with respect to RD and a maximum lattice friction due to precipitation of $\tau_{ppt,max} = -10$ MPa were in good agreement with the experimental results for the AA6014-T4 aluminium alloy, this configuration was further investigated. Figure 8 (a) depicts the RVE used for the following analysis, as well as all the crystal plasticity simulations presented earlier. Parameters for the new modelling approach were identified through a two-step process as outlined in Section 3.3. The results of this parameter identification are illustrated in Fig. 8 (b) in the form of the corresponding stress-strain curves in RD. As was the case with the initial crystal plasticity constitutive model, the results of the new modelling approach match the experimentally determined stress-strain curves in RD with high accuracy.

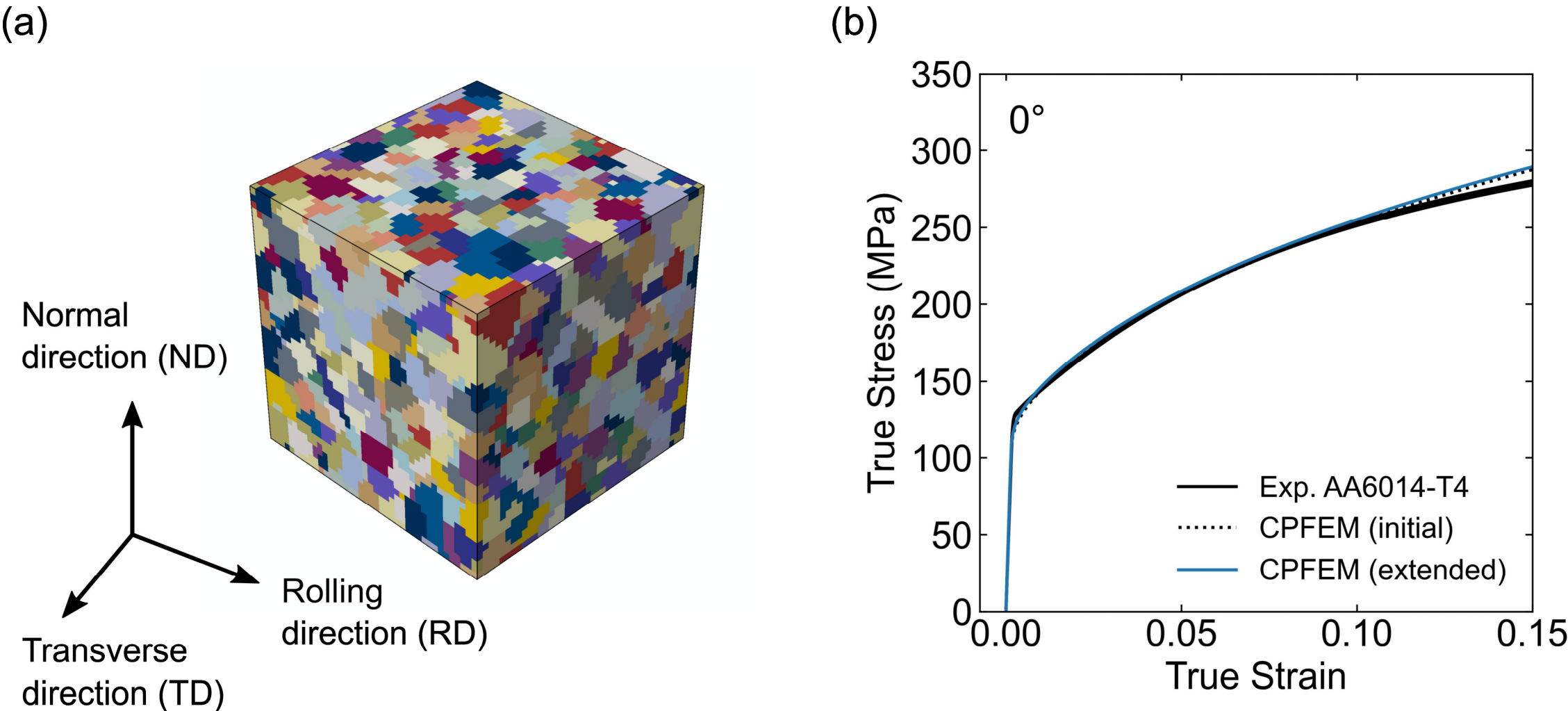

Fig. 8: (a) Representative volume element (RVE) for AA6014-T4. Each set of same-coloured finite elements represents one grain of the microstructure. (b) Stress-strain curves as predicted by the extended crystal plasticity constitutive model compared with the results of the initial crystal plasticity constitutive model serving as a reference, and experimental data.

The experimental stress-strain curves at 15°, 30°, 45°, 60°, 75° and 90° with respect to RD, which were part of the parameter identification as well, are also accurately reproduced by the new modelling approach, as shown in Fig. 9.

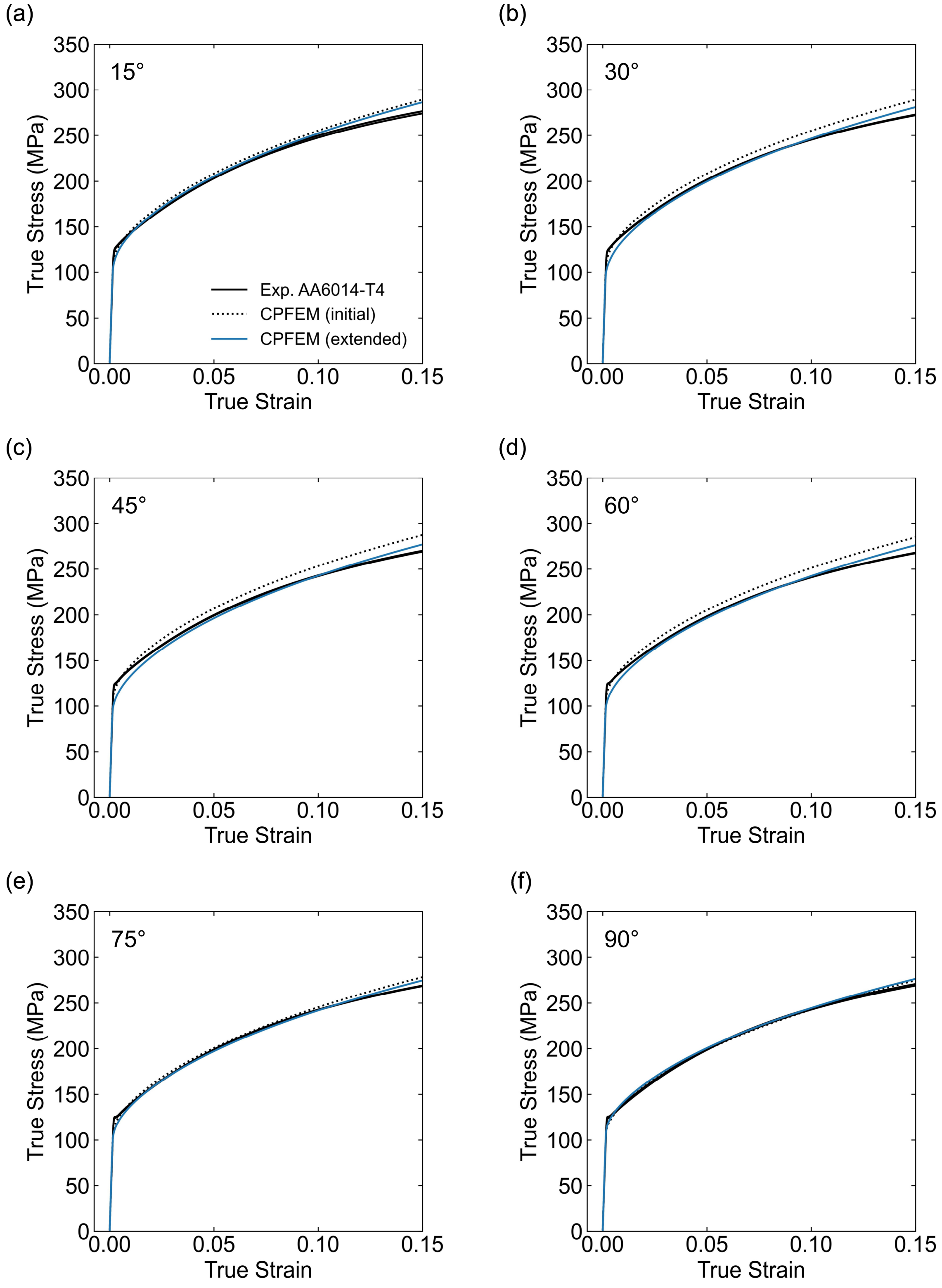

Fig. 9: Stress-strain curves at (a) 15°, (b) 30°, (c) 45°, (d) 60°, (e) 75° and (f) 90° with respect to RD as predicted by the extended crystal plasticity constitutive model compared with results of the initial crystal plasticity constitutive model serving as a reference, as well as experimental data.

In line with the experimental data, normalised yield stresses were determined at a specific plastic work of 15.49 MPa, corresponding to 0.08 true plastic strain, and are illustrated in Fig. 10 (a). While the results of the initial crystal plasticity constitutive model exhibit a rather complementary curve progression as outlined in Section 4.3, the results for the new modelling approach show good agreement with the experimental yield stresses of the AA6014-T4 aluminium alloy. The r-values in Figure 10 (b) were analysed between 0.1 and 0.175 true plastic strain. Again, the results of the extended crystal plasticity constitutive model align well with the experimental data.

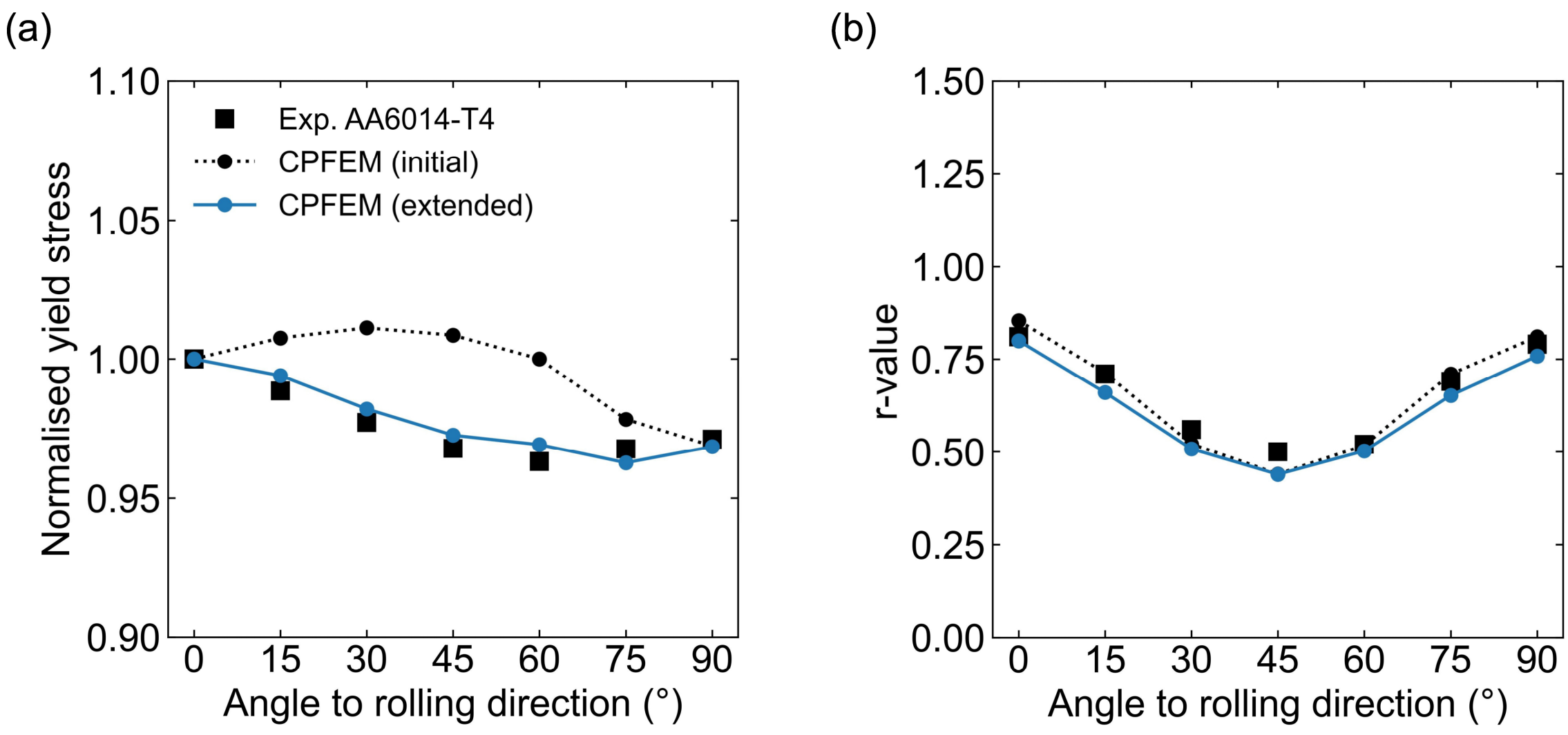

Fig. 10: (a) Normalised yield stresses and (b) r-values predicted by the new modelling approach incorporating precipitation-related effects. Results of the initial crystal plasticity model serving as a reference as well as experimental results for AA6014-T4 aluminium alloy are also shown for comparison.

As a supplement to Fig. 10 (b), Figure 11 illustrates the evolution of the instantaneous r-values for all seven loading directions. As before, the results for the new modelling approach are in good agreement with the experimental r-values of the AA6014-T4 aluminium alloy, though slightly less accurate than the results of the initial crystal plasticity model. All crystal plasticity parameters identified for both the initial and the extended crystal plasticity constitutive models as well as those taken from the literature are summarised in Appendix A.

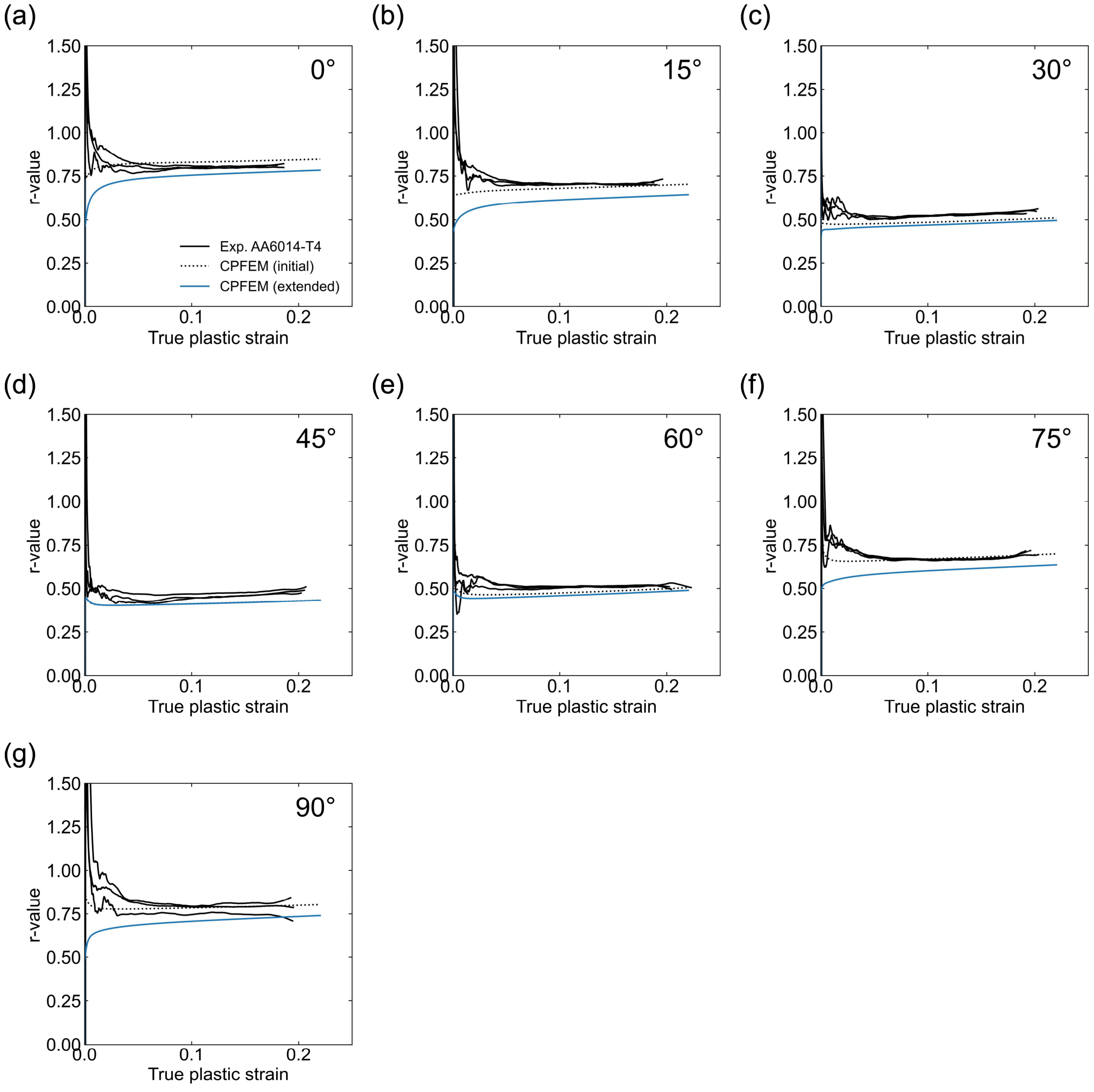

Fig. 11: Evolution of the instantaneous r-values at (a) 0°, (b) 15°, (c) 30°, (d) 45°, (e) 60°, (f) 75° and (g) 90° with respect to RD as obtained from both the extended and the initial crystal plasticity constitutive models compared with the experimental data for AA6014-T4 aluminium alloy.

## 4.6 Anisotropic yield surface

To examine the effect of the new modelling approach on the entire yield surface, both the initial and extended crystal plasticity constitutive models were used to perform 112 crystal plasticity simulations within the plane stress state, as described in Section 3.3. The resulting yield points for the new modelling approach are depicted in Fig. 12 (a) by way of example, showing an even point distribution over one quarter of the yield surface. Each marker represents one point on the yield surface. These points were then used to identify the parameters of the Yld2004-18p yield function. The resulting yield surfaces, as derived for the initial and the extended crystal plasticity constitutive models in Fig. 12 (b), show that the outer shear contours ($\sigma_{12} = 0$) are nearly identical. Differences in the normalised yield surfaces occur primarily at higher contour levels. In this regard, the normalised simple shear stress $\sigma_{12}/\sigma_{0°}$ for Yld2004-18p (CPFEM initial) is 0.62 compared to 0.60 for Yld2004-18p (CPFEM extended). These differences in the normalised yield surface are also evident in Fig. 12 (c), particularly for a uniaxial loading at 45° with respect to RD. The Yld2004-18p yield surface as obtained from simulations using the initial crystal plasticity constitutive model exhibits a complementary curve progression relative to the experimental results. In contrast, the Yld2004-18p yield surface resulting from crystal plasticity simulations using the new modelling approach matches the experimental yield stresses with high accuracy. All parameters identified in this section for the Yld2004-18p yield function are provided in Appendix B.

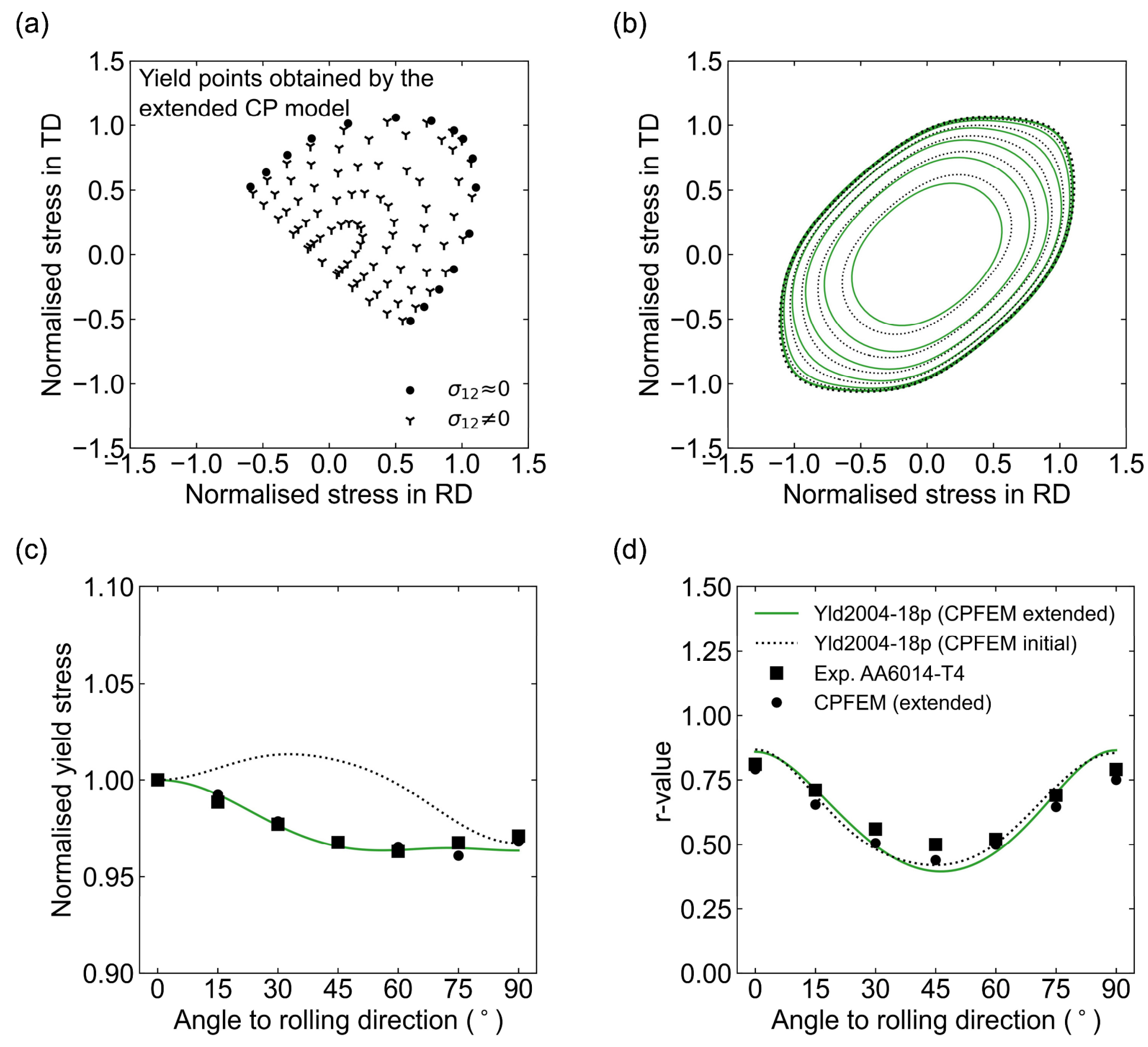

Fig. 12: Yld2004-18p yield surfaces identified from 112 crystal plasticity simulations utilising the initial and the extended crystal constitutive plasticity models: (a) 112 yield points as obtained from the extended crystal plasticity constitutive model using a specific plastic work of 15.49 MPa by way of example, (b) normalised yield surface with respect to the RD-TD plane, (c) normalised yield stresses and (d) r-values with respect to RD. Normalised shear contours are shown in increments of 0.1 from 0.0 to 0.5.

4.7 Non-textured material

As outlined in Section 3.3, the new modelling approach for incorporating precipitation-related effects into crystal plasticity constitutive models was further investigated by examining its effect on a microstructure model with a uniform distribution of crystal orientations. A total of five sets, each containing 1000 Euler angles randomly drawn from a uniform ODF, were applied to the microstructure model shown in Fig. 8 (a). Two exemplary sets of 1000 Euler angles are illustrated in Fig. 13 as ODFs in the reduced Euler space, demonstrating a nearly uniform distribution of crystal orientations.

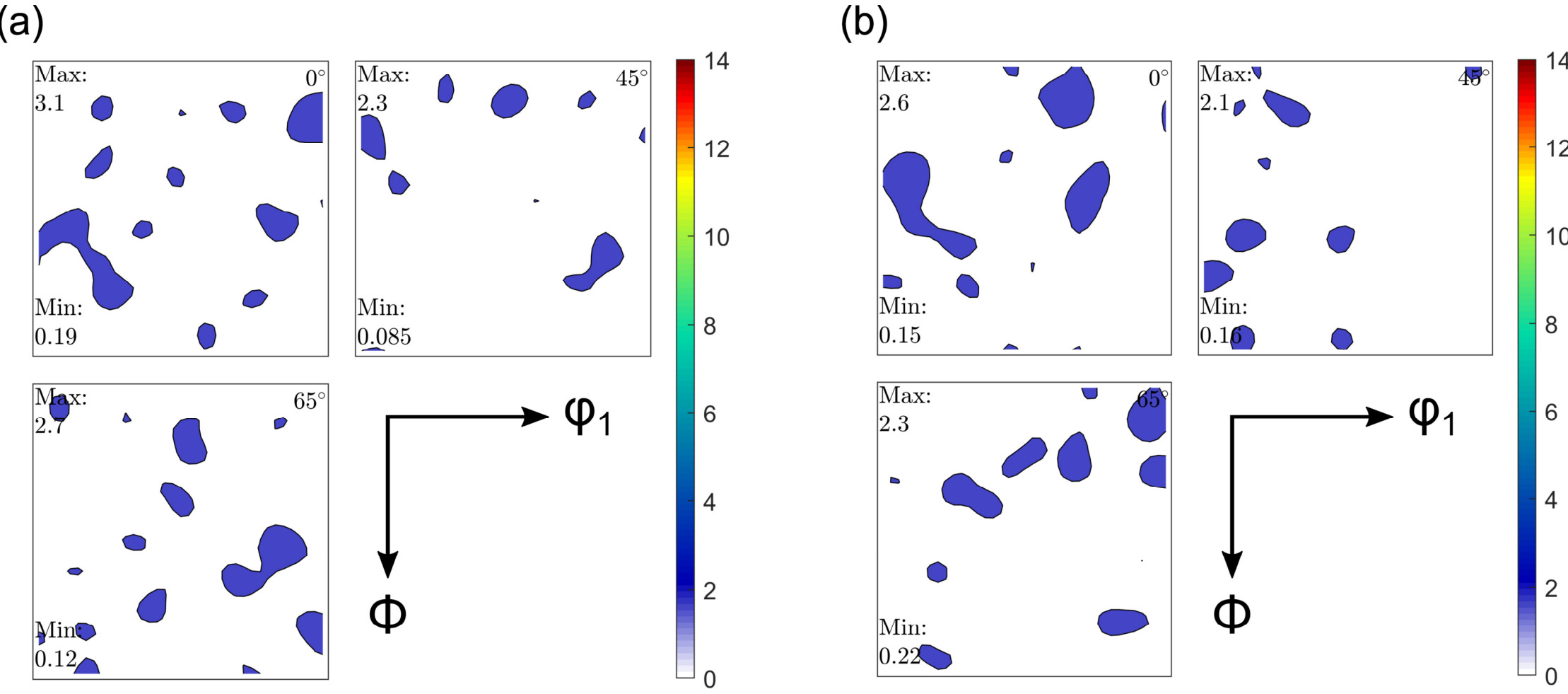

Fig. 13: ODFs for (a) set 1 and (b) set 3 generated from 1000 Euler angles drawn from a uniform ODF and represented as $\varphi_2$-sections at 0°, 45° and 65° through the reduced Euler space.

The results of the crystal plasticity simulations, corresponding to uniaxial tensile tests at 0°, 15°, 30°, 45°, 60°, 75° and 90° with respect to RD for the five sets of uniformly distributed crystal orientations, are shown in Fig. 14. Since a set of 1000 Euler angles with random crystal orientations is still on the low side to represent a non-textured material, i.e. a uniform distribution of crystal orientations, with sufficient statistical confidence, the normalised yield stresses and r-values obtained from the five sets were averaged. As expected, the reference results from the initial crystal plasticity constitutive model exhibit nearly isotropic plastic material behaviour. In contrast, the results for the extended crystal plasticity constitutive model deviate from ideal isotropic plastic material behaviour, as shown in Fig. 14. As with the results in Fig. 10, the new modelling approach leads to a decrease in the normalised yield stresses at 45° with respect to RD compared to the initial crystal plasticity constitutive model. The effect on the r-values is minimal. Overall, the extended crystal plasticity modelling approach causes a slight reduction in the r-values at 0° and 90° with respect to RD.

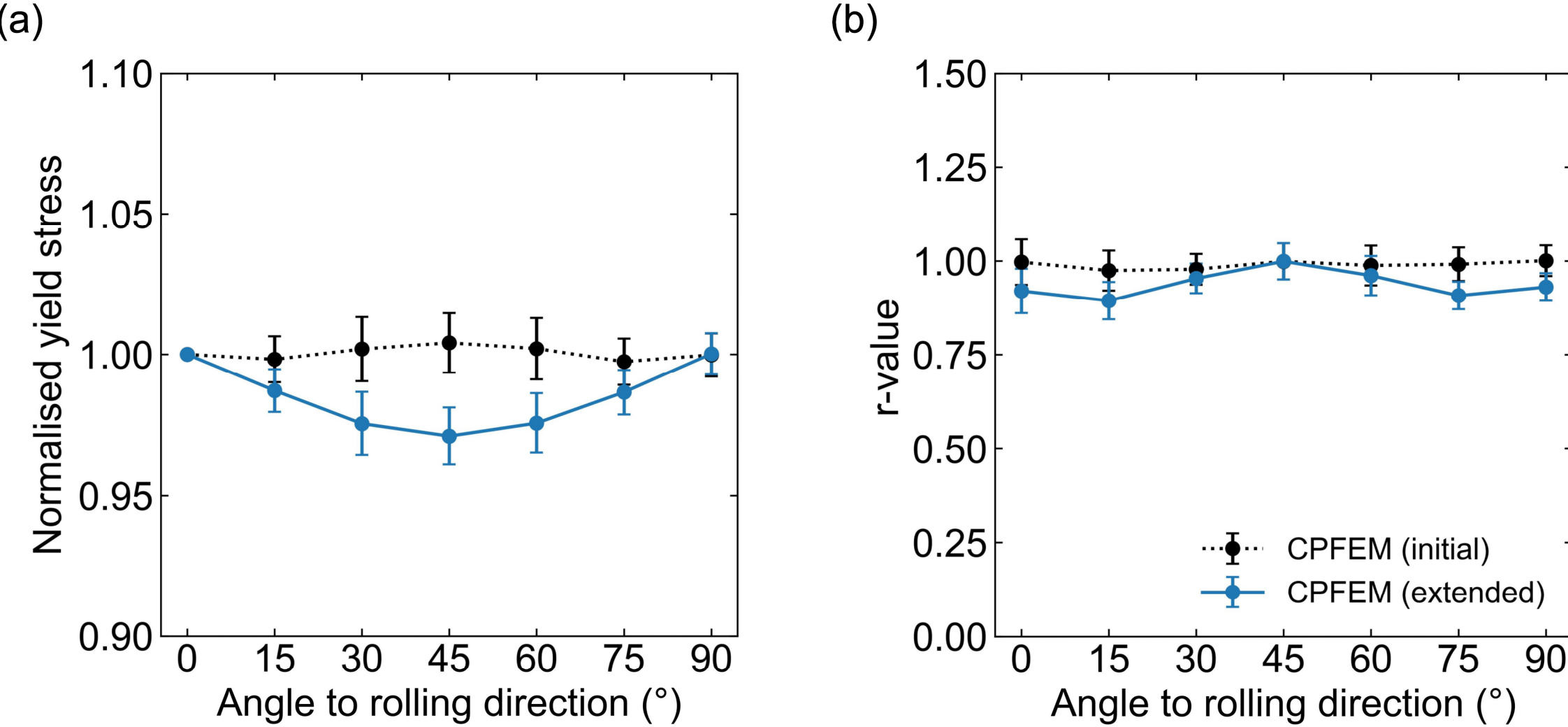

Fig. 14: Averaged (mean ± standard deviation) (a) normalised yield stresses and (b) r-values obtained from crystal plasticity simulations utilising the initial and the extended crystal plasticity constitutive models. For each crystal plasticity constitutive model, five sets containing 1000 randomly drawn Euler angles from a uniform ODF were analysed. All yield stresses were determined based on a specific plastic work of 15.49 MPa, while r-values were analysed between 0.1 and 0.175 true plastic strain.

## 5. Discussion

5.1 Plastic anisotropy in AA6014-T4 aluminium alloy

The experimental results presented in Sections 4.1 and 4.2 demonstrate that the plastic anisotropy of the AA6014 aluminium alloy studied is affected by the temper condition. In the T4 temper, the normalised yield stresses were lowest at 60° with respect to RD, as shown in Fig. 5 (a). After annealing to the O temper, the normalised yield stresses exhibited a complementary curve progression, with the lowest normalised yield stress occurring at 30° with respect to RD. Differences in the r-values between AA6014-T4 and AA6014-O were minimal, with only slight variations at 0° and 90° with respect to RD. Furthermore, the crystallographic textures, as well as the grain size of the AA6014 aluminium alloy in the T4 and O tempers, were almost identical, as shown in Fig. 2 and Fig. 3. These findings are consistent with the literature. As noted in the introduction, similar observations were reported by Kuwabara et al. (2017) for an AA6016 aluminium alloy and by Yoshida et al. (2021) for an unspecified AA6XXX series aluminium alloy. In both studies, a change in plastic anisotropy from the T4 to the O temper was observed, while the crystallographic textures were almost identical. The experimental results for the AA6014 aluminium alloy presented in this study verify the effect of the temper condition on plastic anisotropy for another age-hardenable aluminium alloy of the AA6XXX series.

Regarding the physical mechanisms behind this phenomenon, there is some evidence suggesting that this change in the plastic anisotropy of age-hardenable aluminium alloys is most likely caused by precipitation. First, as mentioned before, it is well documented that precipitation can affect the plastic anisotropy of age-hardenable aluminium alloys, see Hosford and Zeisloft (1972), Jobson and Roberts (1977) and Bate et al. (1981, 1982) for example. More recently, Kuwabara et al. (2017) also linked the change in plastic anisotropy of an AA6016 aluminium alloy from the T4 to the O temper to precipitation, specifically GP zones. Second, although no detailed analysis of precipitation was conducted, the results of the uniaxial tensile tests in Section 4.2 indicate that there is a change in precipitation for the AA6014 aluminium alloy from the T4 to the O temper. While the stress-strain curves for AA6014-T4 in Figure 4 (a) show a yield strength of approximately 240 MPa, the yield strength for AA6014-O decreases to around 110 MPa in Figure 4 (b). Since precipitation is the primary hardening mechanism in age-hardenable aluminium alloys, this change in yield strength can most likely be attributed to a change in precipitation. Third, the results of the crystal plasticity simulations in Fig. 5 and Fig. 10 provide further evidence that the plastic anisotropy of AA6014-T4 is governed by crystallographic texture and precipitation. In contrast, the plastic anisotropy of AA6014-O is controlled by crystallographic texture alone. Based on these three aspects, precipitation is considered the most likely explanation for the change in plastic anisotropy of the AA6014 aluminium alloy. To gain a more comprehensive understanding of the effect of precipitation on texture-induced plastic anisotropy and investigate the underlying physical mechanisms behind it, further experimental analysis employing atomistic techniques, such as transmission electron microscopy (TEM) or atom probe tomography (APT), is necessary.

## 5.2 New modelling approach

Building upon the discussion of plastic anisotropy in AA6014 aluminium alloy in the previous section, the crystal plasticity results in Fig. 5 demonstrate that the initial crystal plasticity constitutive model is well-suited for capturing the effect of texture-induced plastic anisotropy and is thus in good agreement with the experimental results for the AA6014-O aluminium alloy. In contrast, the initial crystal plasticity constitutive model lacks accuracy with respect to the normalised yield stresses for the AA6014-T4 aluminium alloy, as the texture-induced plastic anisotropy is most likely affected by precipitation. Here, the new modelling approach to incorporate precipitation-related effects into conventional crystal plasticity constitutive models appears to be highly effective in improving the accuracy with respect to the normalised yield stresses of the AA6014-T4 aluminium alloy, as shown in Fig. 10. Moreover, the results of the anisotropic yield surfaces in Section 4.6 show that the new modelling approach or rather the extended crystal plasticity constitutive model improves the prediction accuracy of crystal plasticity simulations under uniaxial loading conditions, without causing undesirable distortions in other areas of the yield surface, such as the biaxial or plane strain areas.

Besides its overall suitability, the new modelling approach is advantageous as it relies on only two additional variables: the global preferred direction $\mathbf{d}$ and the maximum lattice friction due to precipitation $\tau_{\mathrm{ppt,max}}$. In Section 4.4, the former has been identified as 45° with respect to RD, which is expected to hold for other AA6XXX series aluminium alloys in the T4 temper. Therefore, the maximum lattice friction due to precipitation $\tau_{\mathrm{ppt,max}}$ is considered the only free parameter. Other crystal plasticity constitutive models which also take precipitation-related effects on the plastic anisotropy into account tend to be more complex. For instance, Mishra et al. (2017) proposed a modified plastic inclusion model, which was implemented into a Taylor model. A comparison with experimental results for an AA6061 aluminium alloy under different temper conditions demonstrated that the modified plastic inclusion model was capable of taking the effect of precipitation-related plastic anisotropy into account. Li et al. (2022) also used a plastic inclusion model to incorporate the effect of precipitation on plastic anisotropy into a crystal plasticity framework. Besides plastic anisotropy, the latter crystal plasticity constitutive model also successfully predicted the yield strength, work hardening and Bauschinger behaviour for an AA6061 aluminium alloy with reasonable accuracy. In contrast to these two crystal plasticity constitutive models, the new modelling approach is relatively simple and easy to implement into most conventional crystal plasticity constitutive models available in the literature.

The new modelling approach is based on the hypothesis that precipitation induces an additional directional dependence with respect to a global material orientation, which superimposes with the texture-induced plastic anisotropy. While a detailed explanation of the underlying physical mechanisms is beyond the scope of this study, this effect could arise from ellipsoidal cluster/GP zones being aligned with a globally preferred direction and/or spherical cluster/GP zones being distributed according to such a preferred direction, as illustrated in Fig. 1. In this context, Hosford and Agrawal (1975) demonstrated that external stresses can affect the orientation of θ' platelets in an Al-Cu single crystal lattice. For example, tensile stresses applied parallel to the [001] crystal direction favoured the formation of θ' platelets on the (001) crystal plane. In contrast, similar formations on the (010) and (100) crystal planes were inhibited. The effect of pressure loading was also studied, specifically the application of compressive stresses parallel to the [001] crystal direction was shown to favour θ' precipitation on (010) and (100) crystal planes and inhibited precipitation on (001) crystal planes. A similar phenomenon may also be plausible for the rolling process of aluminium sheets. Overall, the results of the extended crystal plasticity constitutive model support the claim that there might be a global preferred direction for precipitation in sheet metal, albeit without providing a complete explanation. Again, a more detailed analysis of precipitation is indispensable to provide a more comprehensive understanding of the actual physical mechanisms.

## 6. Conclusions

This study introduces a new modelling approach to incorporate precipitation-related effects on the plastic anisotropy of age-hardenable AA6XXX series aluminium alloys in the T4 temper into crystal plasticity constitutive models. To this end, the effect of the temper condition on the plastic anisotropy was initially analysed for an AA6014 aluminium alloy using uniaxial tensile tests in different directions and EBSD measurements. Subsequently, the new modelling approach was implemented into a conventional crystal plasticity constitutive model and applied to predict the plastic anisotropy of the AA6014-T4 aluminium alloy. In summary, the following conclusions can be drawn from the present work:

(1) Uniaxial tensile tests on AA6014 aluminium alloy in the T4 and O tempers confirm a change in the plastic anisotropy due to the temper condition. Since the crystallographic texture of both AA6014-T4 and AA6014-O was almost identical, this change is most likely caused by precipitation.

(2) The new modelling approach focuses on capturing the overall effect of precipitation on plastic anisotropy in the simplest manner rather than providing a complete physical explanation. As a result, the crystal plasticity constitutive model is extended by two parameters, of which only one is treated as a free parameter.

(3) The capability of the new modelling approach is successfully demonstrated by predicting the plastic anisotropy of the AA6014-T4 aluminium alloy. Compared to a state-of-the-art crystal plasticity constitutive model, the results using the new modelling approach showed better agreement with the experimental data.

**CRediT authorship contribution statement**

**A. Wessel:** Conceptualisation, Methodology, Software, Investigation, Writing – original draft, Visualisation, Funding acquisition. **E. S. Perdahcıoğlu:** Conceptualisation, Methodology, Software, Writing – original draft. **A. H. van den Boogaard:** Conceptualisation, Methodology, Writing – review & editing. **A. Butz:** Writing – review & editing, Supervision, Funding acquisition. **W. Volk:** Writing – review & editing, Supervision.

**Declaration of competing interest**

The authors declare that they have no known competing financial interests or personal relationships that could have appeared to influence the work reported in this paper.

**Data availability**

Data will be made available on request.

**Acknowledgement**

The authors gratefully acknowledge funding from the Federal Ministry for Economic Affairs and Climate Action via the German Federation of Industrial Research Associations – AiF (Arbeitsgemeinschaft industrieller Forschungsvereinigungen e.V.) within the scope of the programme for Industrial Collective Research (Industrielle Gemeinschaftsforschung, IGF), grant number 21466 N. The authors would like to thank Jürgen Timm from Novelis Switzerland SA for providing the AA6014-T4 aluminium sheets. A. Wessel would like to express his gratitude to the Fraunhofer-Gesellschaft for funding his 5-month research stay at the University of Twente via the Fraunhofer International Mobility (FIM) programme. A. Wessel also thanks Valerie Scholes for proofreading this article. Open Access funding enabled and organised by Projekt DEAL.

## Appendix A. Crystal plasticity parameters

Parameters identified for both the initial and the extended crystal plasticity constitutive models representing the AA6014-T4 aluminium alloy are summarised in Tables A.1 and A.2, respectively.

Table A.1: Parameters of the initial crystal plasticity model as representative for AA6014-T4 aluminium alloy.

| Parameter | Description | Unit | Value | Reference |
| --- | --- | --- | --- | --- |
| $C_{11}$ | Elastic constant | MPa | 106750 | Haynes (2014) |
| $C_{12}$ | Elastic constant | MPa | 60410 | Haynes (2014) |
| $C_{44}$ | Elastic constant | MPa | 28340 | Haynes (2014) |
| $b$ | Burgers vector | mm | 2.86E-07 | Perdahcıoğlu et al. (2018), Hansen and Huang (1998) |
| $\tau$ | Lattice friction | MPa | 18.0 | [a] |
| $\rho_0$ | Initial dislocation density | $mm^{-2}$ | 1.00E7 | [a] |
| $\rho_s$ | Saturation dislocation density | $mm^{-2}$ | 6.98E8 | [a] |
| $\gamma_s$ | Parameter controlling the saturation rate | - | 0.21 | [a] |
| $q_0$ | Self-interaction | - | 0.122 | Kubin et al. (2008) |
| $q_1$ | Coplanar interaction | - | 0.122 | Kubin et al. (2008) |
| $q_2$ | Collinear interaction | - | 0.625 | Kubin et al. (2008) |
| $q_3$ | Orthogonal interaction | - | 0.070 | Kubin et al. (2008) |
| $q_4$ | Gissile interaction | - | 0.137 | Kubin et al. (2008) |
| $q_5$ | Sessile interaction | - | 0.122 | Kubin et al. (2008) |

[a] Identified by a reverse engineering approach

Table A.2: Parameters of the extended crystal plasticity model utilising the new modelling approach for incorporating precipitation-related effects. Parameters are representative for AA6014-T4 aluminium alloy.

| Parameter | Description | Unit | Value | Reference |
|---|---|---|---|---|
| $C_{11}$ | Elastic constant | MPa | 106750 | Haynes (2014) |
| $C_{12}$ | Elastic constant | MPa | 60410 | Haynes (2014) |
| $C_{44}$ | Elastic constant | MPa | 28340 | Haynes (2014) |
| $b$ | Burgers vector | mm | 2.86E-07 | Perdahcioğlu et al. (2018), Hansen and Huang (1998) |
| $\tau_{\text{ini}}$ | Lattice friction | MPa | 23.5 | [a] |
| $\tau_{\text{ppt,max}}$ | Lattice friction due to precipitation | MPa | 12.5 | [a] |
| $\rho_0$ | Initial dislocation density | $mm^{-2}$ | 1.00E7 | [a] |
| $\rho_{\text{s}}$ | Saturation dislocation density | $mm^{-2}$ | 6.98E8 | [a] |
| $\gamma_{\text{s}}$ | Parameter controlling the saturation rate | - | 0.21 | [a] |
| $q_0$ | Self-interaction | - | 0.122 | Kubin et al. (2008) |
| $q_1$ | Coplanar interaction | - | 0.122 | Kubin et al. (2008) |
| $q_2$ | Collinear interaction | - | 0.625 | Kubin et al. (2008) |
| $q_3$ | Orthogonal interaction | - | 0.070 | Kubin et al. (2008) |
| $q_4$ | Gissile interaction | - | 0.137 | Kubin et al. (2008) |
| $q_5$ | Sessile interaction | - | 0.122 | Kubin et al. (2008) |

[a] Identified by a reverse engineering approach

## Appendix B. Yield function parameters

Parameters of the Yld2004-18p yield function, as identified from 112 crystal plasticity simulations for the AA6014-T4 aluminium alloy, are provided in Table B1. Crystal plasticity simulations were performed using both the initial and the extended crystal plasticity constitutive models.

Table B.1: Parameters of the Yld2004-18p yield function as identified from 112 crystal plasticity simulations using the initial and the extended crystal plasticity constitutive model. Parameters related to out-of-plane anisotropy were not considered in the analysis and were set to their isotropic values.

| | Yld2004-18p (CPFEM initial) | Yld2004-18p (CPFEM extended) |
|---|---|---|
| $c'_{12}$ | 1.0 | 1.0 |
| $c'_{13}$ | 1.0 | 1.0 |
| $c'_{21}$ | 0.3281 | 0.8635 |
| $c'_{23}$ | 1.6027 | 0.6438 |
| $c'_{31}$ | 1.0117 | 0.7826 |
| $c'_{32}$ | 0.9462 | 0.8065 |
| $c'_{44}$ | 1.0 | 1.0 |
| $c'_{55}$ | 1.0 | 1.0 |
| $c'_{66}$ | 0.6295 | 0.6525 |
| $c''_{12}$ | 1.0698 | 0.9243 |
| $c''_{13}$ | 1.0323 | 1.3559 |
| $c''_{21}$ | 1.0837 | 0.9887 |
| $c''_{23}$ | 1.2525 | 1.4399 |
| $c''_{31}$ | 0.5165 | -0.3709 |
| $c''_{32}$ | -0.1314 | 0.4356 |
| $c''_{44}$ | 1.0 | 1.0 |
| $c''_{55}$ | 1.0 | 1.0 |
| $c''_{66}$ | 1.1326 | 1.1607 |